\documentclass[
reprint,
superscriptaddress,
amsmath,amssymb,
aps,prl
]{revtex4-2}

\usepackage{amsfonts}

\usepackage{graphicx}
\usepackage{dcolumn}
\usepackage{bm}
\usepackage{hyperref}
\usepackage{physics}
\usepackage{xcolor}
\usepackage{mathrsfs}
\usepackage{mathtools}

\newcommand{\mean}[1]{\left\langle#1\right\rangle}
\newcommand{\X}{\sigma^x}

\newcommand{\Z}{\sigma^z}

\begin{document}

%
%
\title{Work statistics, quantum signatures and enhanced work extraction in quadratic fermionic models}


\author{Alessandro Santini}
\email{asantini@sissa.it}
\affiliation{SISSA, via Bonomea 265, 34136 Trieste, Italy}
 
\author{Andrea Solfanelli}
\email{asolfane@sissa.it}
\affiliation{SISSA, via Bonomea 265, 34136 Trieste, Italy}
\affiliation{INFN, via Bonomea 265, 34136 Trieste, Italy}
\affiliation{Center for Life Center for Life Nano-Neuro Science @ La Sapienza, Italian Institute of Technology, 00161 Roma, Italy}

\author{Stefano Gherardini}
\email{stefano.gherardini@ino.cnr.it}
\affiliation{CNR-INO, Area Science Park, Basovizza, 34149 Trieste, Italy}
\affiliation{SISSA, via Bonomea 265, 34136 Trieste, Italy}
\affiliation{
ICTP, Strada Costiera 11, 34151 Trieste, Italy}

\author{Mario Collura}
\email{mcollura@sissa.it}
\affiliation{SISSA, via Bonomea 265, 34136 Trieste, Italy}
\affiliation{INFN, via Bonomea 265, 34136 Trieste, Italy}

\date{\today}

\begin{abstract}
In quadratic fermionic models we determine a quantum correction to the work statistics after a sudden and a time-dependent driving. Such a correction lies in the non-commutativity of the initial quantum state and the time-dependent Hamiltonian, and is revealed via the Kirkwood-Dirac quasiprobability (KDQ) approach to two-times correlators. Thanks to the latter, one can assess the onset of non-classical signatures in the KDQ distribution of work, in the form of negative and complex values that no classical theory can reveal. By applying these concepts on the one-dimensional transverse-field Ising model, we relate non-classical behaviours of the KDQ statistics of work in correspondence of the critical points of the model. Finally, we also prove the enhancement of the extracted work in non-classical regimes where the non-commutativity takes a role.
\end{abstract}

\maketitle



A fundamental quantity in non-equilibrium thermodynamics is the statistics of the work done on a quantum system by an external coherent source varying its Hamiltonian over time~\cite{EspositoRMP2009,CampisiRMP2011,AllahverdyanPRE2014}. In the quantum regime, the task of determining energy-change fluctuations is still debated in the current literature, especially when considering many-body quantum systems~\cite{Silva2008PRL,Foini2011PRB,Marino2014PRB,Goold2018,Zhaoyu2019PRR,DelCampo2020PRR,Zhaoyu2020PRL} and addressing cases where the initial density matrix $\rho_0$ and the system Hamiltonian $H_t$ are non-commuting operators~\cite{Llobet2015PRX,LevyPRXQuantum2020,MaffeiPRR2021,SolinasPRA2022,LostaglioArXiv2022,Hernandez-GomezArXiv2022,MaffeiArXiv2022}. 

It is known that there is no quantum observable that allow us to measure directly differences of energy values in different realizations of the system dynamics~\cite{TalknerPRE2007}. Hence, different protocols for their evaluation have been proposed in the last few decades~\cite{DeffnerPRE2016,Paternostro2020Entropy,SonePRL2020,MicadeiPRL2020,MicadeiPRL2021,SolinasPRA2021,SolinasPRA2022,GherardiniPRA2021,Hernandez-GomezArXiv2022coherence,LostaglioArXiv2022}. In this regard, a celebrated protocol is the two-point measurement (TPM) scheme~\cite{Kurchan2000,CampisiPRL2009,KafriPRA2012,Hernandez-GomezPRR2020} that well reproduces the quantum work statistics when $\rho_0$ and $H_t$ commute. However, in case of non-commutativity, the results provided by the TPM scheme suffer of the uncertainty due to quantum measurement back-action. On the other hand, in agreement with the no-go theorems in Refs.~\cite{Perarnau-LlobetPRL2017,HovhannisyanArXiv2021,LostaglioArXiv2022}, it is also known that there is not a unique measurement scheme to characterize---quantum mechanically---the work statistics or, more in general, functions of measurement outcomes defined at two times.

In this paper, we compute the characteristic function of the work distribution for a quadratic fermionic many-body system~\cite{barouch_first,*barouch_second,*barouch_third}, using the Kirkwood-Dirac quasiprobability (KDQ) approach~\cite{kirkwood1933quantum,dirac1945analogy,YungerPRA2018,Lupu-GladsteinPRL2022} applied to quantum thermodynamics~\cite{YungerPRA2017,LostaglioArXiv2022,Hernandez-GomezArXiv2022}. In fact, quadratic models are of paramount importance since they allow to investigate equilibrium and non-equilibrium properties of probability distribution functions exactly~\cite{Ivanov_2013,Klich_2014,Rajabpour_2017,Groha_2018,Collura_2019}. They give us the opportunity to rigorously inspect at quantum many-particle phenomena, thus going beyond the few-particle results, but still keeping manageable the complexity of the computation.

Thanks to KDQs, we are able to determine a quantum correction to the distribution of work, by thus amending the result obtained from applying the TPM scheme, which rids off any quantum coherence of $\rho_0$ in the initial Hamiltonian basis. In this way, we recover the unperturbed expression of the average work, i.e., 
%
%
$\Tr{H_{t}\rho_t} - \Tr{H_{0}\rho_0}$, which exhibits a natural classical-quantum correspondence~\cite{JarzynskiPRX2015}. Another key property of the KDQ approach is that, by evaluating the characteristic function, 
%
%
one can determine whether the work statistics is originated by a quasiprobability distribution, thus with negative real terms or even complex one. The latter cannot be reproduced by any corresponding classical theories, and for this reason their presence represents a \emph{signature of non-classicality}~\cite{Arvidsson-ShukurJPA2021}. A work distribution can be non-classical at a given time if the initial state and the system Hamiltonian do \emph{not} commute. In such a case, a quantum correction to the work statistics needs to be applied. Here, these concepts are derived for a generic many-body fermionic system, and directly linked to the critical behaviour of the one-dimensional (1D) transverse field Ising model~\cite{sachdev_2000} across its magnetic phases. In this way, we determine that the KDQ distribution of the stochastic work becomes non-classical across the critical point, 
%
%
where the non-commutativity of $\rho_0$ and $H_t$, if present, is made evident. We also find-out 
%
%
a work extraction enhancement in those regions where non-commutativity takes a role. As an important remark, we stress that the results of our analysis hold both in the case the initial Hamiltonian $H_{t_1}$ of the work protocol is powered by a sudden quench, and in the more realistic scenario the Hamiltonian change $H_{t_1} \rightarrow H_{t_2}$ is enabled by a sufficiently fast ramp driving. 

\textbf{Protocol.}---Let us take into account a quantum system starting from a state $\rho_0$ subjected to a time-dependent driving of its Hamiltonian parameters from time $t_1$ to $t_2$. By expressing the Hamiltonian $H_t$ in spectral decomposition, i.e., 
%
%
$H_t = \sum_n E_t^{(n)} \Pi_t^{(n)}$ with $\Pi_t^{(n)} = \dyad*{E_t^{(n)}}$, the probability to do/extract a given amount of stochastic work $W_{[t_1,t_2]} = E_{t_2}-E_{t_1}$ on/from the system obeys the distribution
\begin{equation}
    P(W_{[t_1,t_2]}) = \sum_{m,n} p_{m,n} \, \delta\left( W - (E_{t_2}^{(m)}-E_{t_1}^{(n)}) \right)
\end{equation}
where $\delta(\cdot)$ is the Kronecker delta, and $E_{t_1}^{(n)}$ and $E_{t_2}^{(m)}$ are, respectively, the internal energies of the system at the beginning and at the end of the driving. In fact, since the energy of the system is assumed to change according to a time-dependent driving of its Hamiltonian, the internal energy variations can be ascribed as work. Moreover, with $p_{m,n}$ we denote the 
%
%
joint probability of the energy at times $t_1$ and $t_2$. As argued in the introduction, if $\comm{\rho_0}{H_{t_1}}\neq 0$ and/or $\comm{H_{t_1}}{H_{t_2}}\neq 0$ there is not an unique way of defining multi-times joint probabilities in the quantum regime. In fact, quantum mechanically, one cannot obtain the statistics of outcomes originating from non-compatible quantum observable without loosing information~\cite{Perarnau-LlobetPRL2017,HovhannisyanArXiv2021,LostaglioArXiv2022}. Therefore, we are going to analyze the following two possibilities:
\begin{equation}\label{eq:joint_probs}
    p_{m,n} = 
    \begin{dcases}
    \Tr{U_{[t_1,t_2]}\Pi_{t_1}^{(n)}\rho_0 \, \Pi_{t_1}^{(n)} U^\dagger_{[t_1,t_2]} \Pi_{t_2}^{(m)}}& \mathrm{TPM} \\
    %
    %
    \Tr{U_{[t_1,t_2]}\Pi_{t_1}^{(n)}\rho_0 \, U^\dagger_{[t_1,t_2]} \Pi_{t_2}^{(m)}}& \mathrm{KDQ}
    \end{dcases}
\end{equation}
where 
%
%
$U_{[t_1,t_2]}=\mathcal{T}{\rm exp}(-i\int_{t_1}^{t_2} H_t dt)$ is the time-ordered exponential of the Hamiltonian. The expressions in (\ref{eq:joint_probs}) are the joint probabilities returned by the TPM and KDQ approach, respectively, to the two-times work statistics. It is worth noting that, if $\comm{\rho_0}{H_{t_1}}=0$, then the two schemes are equivalent. 
Otherwise, the non-commutativity of $\rho_0$ and $H_{t_1}$ can entail that some $p_{m,n}$ are negative real numbers or even complex. In this context, the characteristic function $G(u)=\mean{e^{iuW_{[t_1,t_2]}}}$ of the work distribution is thus defined as
\begin{align}
    G(u) & = \begin{dcases}
    \Tr{\Delta_1(\rho_{0}) e^{-iuH_{t_1}} e^{iuH^\mathrm{H}_{t_2}} } & \mathrm{TPM}\\
    \Tr{\rho_{0} \, e^{-iuH_{t_1}}e^{iuH^\mathrm{H}_{t_2}} } & \mathrm{KDQ}
    \end{dcases}
\end{align}
%
%
where $u\in\mathbb{C}$, $H_{t_2}^\mathrm{H} = U^\dagger_{[t_1,t_2]} H_{t_2} U_{[t_1,t_2]}$ denotes the evolution of $H_{t_2}$ expressed in Heisenberg representation, and $\Delta_1(\rho_{0})= \sum_{n}\Pi_{t_1}^{(n)}\rho_{0}\Pi_{t_1}^{(n)}$ is the diagonal part of the initial density matrix $\rho_0$ in the basis of $H_{t_1}$. Note that the characteristic function of the KDQ work distribution is the \emph{quantum correlation function} of the operators $e^{-iuH_{t_1}}$ and $e^{iuH^\mathrm{H}_{t_2}}$ that, in the general case, do not commute among them and with $\rho_0$.
%

\textbf{Model.}---We consider quadratic fermionic models with Hamiltonian  
\begin{equation}
H_t = - \sum_{k,j}\left(T_{kj}c_{k}^{\dagger}c_{j} + \Delta_{kj}c_{k}^{\dagger}c_{j}^{\dagger} + h.c.\right) - h(t)\sum_{k}(2n_k-1)   
\end{equation}
where $c_{k}^{\dagger}$ and $c_{k}$ are the fermionic creation and annihilation operators, such that 
%
%
$n_k = c_{k}^{\dagger}c_{k}$ and 
$\{c_{k},c^{\dag}_{j}\} = \delta_{kj}$.
Moreover, $T_{kj}$ and $\Delta_{kj}$ are respectively the hopping (or tunnelling) and pairing amplitudes, while $h(t)$ time-dependent strength of an external field. Under the assumption that the Hamiltonian $H_t$ is translational invariant, i.e., $T_{kj}=T(|k-j|)$ and $\Delta_{kj}=\Delta(|k-j|)$, $H_t$ admits the quadratic form
%
%
$H_{t} = \sum_{p > 0} \Psi_p^\dagger \mathbb{H}_p(t) \Psi_p$ in the momentum component $p$, with $\Psi_p^{\dag} = ( \tilde{c}_p^{\dag}, \tilde{c}_{-p} )$ where 
\begin{equation*}
    \tilde{c}_p = \frac{e^{i\frac{\pi}{4}}}{\sqrt{L}}\sum_{k}e^{-ipk}c_{k}, \quad
    \mathbb{H}_p(t) = 
    \begin{pmatrix*}[c]
    h(t) - \tilde{T}_{p} & \tilde{\Delta}_{p} \\
    \tilde{\Delta}_{p}  & \tilde{T}_{p} - h(t)
    \end{pmatrix*},
\end{equation*}
in terms of the Fourier transform $\tilde{T}_{p} = \sum_{r}T(r)\cos(pr)$ and $\tilde{\Delta}_{p} = \sum_{r}\Delta(r)\sin(pr)$ of the hopping and pairing amplitudes respectivelly. 
Quadratic fermionic models 
%
%
are quite versatile, since they can be directly mapped to quantum spin systems via the Jordan-Wigner transformation~\cite{Jordan1928berDP}. In this regard, both the transverse field Ising model and the XY model can be recovered and then analytically solved~\cite{Fagotti_RDM_2013}. 

The $2\times 2$ Hamiltonian $\mathbb{H}_p(t)$ becomes diagonal after a proper SU(2) rotation around the $y$ axis. Namely, 
$\mathbb{H}_p(t) = R_{y}^{\dagger}(\phi_p(t))\mathbb{D}_p(t)R_{y}(\phi_p(t))$,
with $\mathbb{D}_p(t) = \omega_p(t)\sigma^z$ and
$R_{y}^{\dagger}(\phi_p(t)) = {\rm exp}(-i\phi_p(t)\sigma^{y}/2)$ where $\phi_p(t)$ denotes the `Bogoliubov angle', and $\sigma^{\alpha}$ ($\alpha \in \{x,y,z\}$) are Pauli matrices.
%
%
We thus get the instantaneous `Bogoliubov fermions' $\Gamma_p(t) = (\gamma_p, \gamma^\dagger_{-p})^T = R_{y}^{\dagger}(\phi_p(t))\Psi_p$ such that, for any time $t$, 
$H_t = \sum_{p>0} \Gamma_p^\dagger(t) \mathbb{D}_p(t)\Gamma_p(t)$.

During the work protocol, we vary the external field over time according to a specific time-dependent function $h(t)$. In the characteristic function 
%
%
$G(u)$, the full time-dependence enters via the Heisenberg representation of the final Hamiltonian $H_{t_2}^\mathrm{H}$, whose closed-form expression reads
\begin{equation}\label{eq:H_t2_Heisenberg}
H_{t_2}^\mathrm{H}  = \sum_{p>0} \Psi^\dagger_p 
{\mathbb{H}^{[2]}_p}^\mathrm{H}
\Psi_p \,, 
\end{equation}
with ${\mathbb{H}^{[2]}_p}^\mathrm{H} = \mathbb{U}_{p,t_1:t_2}^\dagger R^\dagger_y(\phi_p^{[2]}) \omega_p^{[2]}\sigma^z R_y(\phi_p^{[2]})\mathbb{U}_{p,t_1:t_2}$, where $\mathbb{U}_{p,t_1:t_2} = \mathcal{T}{\rm exp}(-i \int_{t_1}^{t_2} \mathbb{H}_{p}(s)\,ds)$, and we used the shortcut notation $\omega_p^{[j]} = \omega_p(t_j)$ and $\phi_p^{[j]} = \phi_p(t_j)$ for the sake of presentation. Note that the time-ordered exponential entering the matrix $\mathbb{U}_{p,t_1:t_2}$ is carried out from solving the Heisenberg differential equation $\dot{\Psi}_{p} = i\comm{H_t}{\Psi_p} = -i\mathbb{H}_p(t)\Psi_p$, with $\hbar$ set to $1$.



\textbf{KDQ characteristic function of work.}---Let us show how to analytically compute the characteristic function of the KDQ work distribution, as well as its derivatives, for a generic quadratic fermionic model. We thus recall that $G(u)$ depends on the initial density matrix that is taken equal to $\rho_0 = \exp{-\beta H_{t_0}}/{Z}$, with $Z=\Tr{e^{-\beta H_{t_0}}}$ such that $\comm{H_{t_0}}{H_{t_1}}\neq 0$. By substituting $\rho_0$, $H_{t_1}$ and $H_{t_2}^{H}$ in the characteristic function, 
%
%
one gets $G(u)=\prod_p g_p(u)/g_p(0)$ where
\begin{equation}\label{eq:gen_expression_charac_function}
    g_p(u) =
    \prod_{p>0} \Tr{e^{-\beta\Psi_p^\dagger \mathbb{H}^{[0]}_p\Psi_p} e^{-iu\Psi_p^\dagger \mathbb{H}^{[1]}_p \Psi_p}  e^{iu\Psi_p^\dagger {\mathbb{H}^{[2]}_p}^\mathrm{H}
 \Psi_p} } \,.
\end{equation}
%
%
We note that the trace in (\ref{eq:gen_expression_charac_function}) is evaluated in the $p$-momentum Fock subspace $\{\ket{\emptyset},c^\dagger_p\ket{\emptyset},c^\dagger_{-p}\ket{\emptyset},c^\dagger_pc^\dagger_{-p}\ket{\emptyset}\}$, where $\ket{\emptyset}$ denotes the vacuum state such that $c_{\pm p}\ket{\emptyset} = 0$. By repeatedly applying the group composition law of SU(2) matrices to Eq.~(\ref{eq:gen_expression_charac_function}), one gets
\begin{equation}
    e^{-\beta\Psi_p^\dagger \mathbb{H}^{[0]}_p\Psi_p} e^{-iu\Psi_p^\dagger \mathbb{H}^{[1]}_p \Psi_p}  e^{iu\Psi_p^\dagger {\mathbb{H}^{[2]}_p}^\mathrm{H}
 \Psi_p} = e^{\Psi_p^\dagger \mathbb{B}_p \Psi_p}.
\end{equation} 
The spectral decomposition of $\mathbb{B}_p$ provides us eigenvalues $b_p,-b_p$ such that
\begin{equation}
    g_p(u) = \prod_{p>0} \Tr{e^{\Psi^\dagger_p\mathbb{B}_p\Psi_p}}=\prod_{p>0} 2(1+\cosh b_p).
\end{equation}

\textbf{Quantum signatures in work statistics.}---The KDQ distribution of work $P(W_{[t_1,t_2]})$ can exhibit non-classical properties, which no classical model can reproduce. Specifically, the non-classicality of $P(W_{[t_1,t_2]})$ means 
%
%
that $\Re{p_{m,n}}<0$ and/or $\Im{p_{m,n}} \neq 0$ for some indices $(m,n)$. To witness 
%
%
$\Im{p_{m,n}} \neq 0$, one can use the following statement:
\begin{equation}
    G^*(u) = \int dW_{[t_1,t_2]} e^{-iuW_{[t_1,t_2]}}P^*(W_{[t_1,t_2]}) = G(-u)
\end{equation}
\emph{if and only if} $\Im{p_{m,n}}=0$ for any $(m,n)$. Hence, a violation of the identity $G^*(u)=G(-u)$ is directly linked to the presence of complex values in the KDQ distribution of work. In the previous paragraph, we have determined that $G(u) = \prod_p g_p(u)/g_p(0)$. As a result, $G^*(u)=G(-u)$ if $g_p(u)=g^*_p(-u)$ for any $p,u$. 
%
%
In this regard, 
\begin{eqnarray}\label{eq:non-classicality_measure}
    g_p(u)-g^*_p(-u) = -4 \sin(\phi^{[0]}_p-\phi^{[1]}_p) \sin(\phi^{[2]}_p)\times\nonumber \\
    \sinh(\beta  \omega^{[0]}_p)(\Im s_p^2+\Im z_p^2)\sin(u \omega^{[1]}_p) \sin(u \omega^{[2]}_p)
\end{eqnarray}
%
%
where $z_p$ and $s_p$ are the independent variables that define the unitary $2\times 2$ matrix $\mathbb{U}_{p,t_1:t_2}$ associated to the $p$-mode:
$\mathbb{U}_{p,t_1:t_2} = \begin{pmatrix*}
        z_p && -s^*_p \\ s_p && z^*_p
\end{pmatrix*}$. The right-hand-side (r.h.s.) of Eq.~(\ref{eq:non-classicality_measure}), witness of non-classicality, is equal to zero if $\comm{\rho_0}{H_{t_1}}=0$, i.e., if $\phi^{[0]}_p=\phi^{[1]}_p$ $\forall p$. Moreover, in the sudden quench limit, 
namely when $H^{H}_{t_2} = H_{t_2}$,
$z_p=1$ and $s_p=0$; therefore, the KDQ distribution of work $P(W_{[t_1,t_2]})$ is a real-valued function of real variable. Nonetheless, we can find quantum signatures due to negative quasiprobabilities. This is shown in detail in the Supplemental Material (SM), where the sign of the $4$-th central moment $\langle(W_{[t_1,t_2]}-\langle W_{[t_1,t_2]}\rangle)^4\rangle$ is investigated for paradigmatic case-studies.

Albeit non-classical, the statistical moments of the KDQ distribution of work can be still computed from making the derivatives of the characteristic with respect to $u$.  Specifically, for the average work, one gets
\begin{equation}\label{eq:mean W}
    \mean{W} = 
    L\int_{0}^{\pi}\frac{dp}{2\pi}\tanh \left(\frac{\beta\omega ^{[0]}_p}{2}\right) \left(\omega^{[1]}_p Q^{[0,1]}_p-\omega^{[2]}_p
    Q^{[0,2]}_p\right),
\end{equation}
where $Q^{[0,1]}_p = 2P_p^{[0,1]}-1 = \cos{(\phi^{[0]}_p-\phi^{[1]}_p)}$ and $Q^{[0,2]}_p = 2P_p^{[0,2]}-1$, with $1-P_p^{[\ell,j]}$ denoting the \emph{transition probability} from the energy eigenstates at time $t_{\ell}$ towards the ones at time $t_j$ in the momentum domain (see SM for the derivation). Notice that in Eq.~(\ref{eq:mean W}), $\mean{W_{[t_1,t_2]}}/L =: \mean{w}$ becomes the \emph{work density} in the thermodynamic limit of $L\to\infty$, with $L$ size of the fermionic system. This result generalizes the findings of Refs.~\cite{Zhaoyu2019PRR,Zhaoyu2020PRL}, where the work protocol operates from an initial state that commutes with the initial Hamiltonian, and the average work is computed using the TPM scheme. In Eq.~(\ref{eq:mean W}), the TPM result is retrieved when $Q_p^{[0,1]}=1$ $\forall p$.

Eq.~(\ref{eq:mean W}) can be applied to any initial state of the form $\rho_0 = e^{-\beta H_{t_0}}/Z_0$ that does not commute in general with the Hamiltonian $H_{t_1}$ at the beginning of the work protocol. As a consequence, also quantum coherences in the initial energy basis start playing a relevant role in energy fluctuations. Their effect on the average work is encoded in the parameters $Q_p^{[0,1]}$ and $Q_p^{[0,2]}$. In particular, $Q_p^{[0,1]}$ is related to the overlap between the eigenbases of the initial Hamiltonian and the initial state respectively. Interestingly, the average work extracted by the external driving, $-\mean{W_{[t_1,t_2]}}$, increases as $Q_p^{[0,1]}$ approaches the minimum value $-1$ for any $p$. $Q_p^{[0,1]}=-1$ corresponds to $\phi^{[0]}_p-\phi^{[1]}_p = \pi$, meaning that the eigenbasis of $\mathbb{H}_p^{[0]}$ and $\mathbb{H}_p^{[1]}$ are orthogonal. In other terms, concerning $Q_p^{[0,1]}$, the work extraction is optimized when the operators $\rho_0$ and $H_{t_1}$ are maximally non-commuting. On the other hand, the parameters $Q_p^{[0,2]}$ are associated to the transition probabilities among the instantaneous eigenstates of the (time-dependent) Hamiltonian $H_t$ from time $t_0$ to $t_2$ as an effect of the external driving. If the transitions operated by the driving field occur in a \emph{non-adiabatic} fashion, then part of the internal energy variation is converted in \emph{irreversible work}~\cite{DornerPRL2012,RevathyPRR2020,Solfanelli2020PRB,SolfanelliArXiv2022}. Accordingly, in the general case, the $Q_p^{[0,2]}$s describe on average how the presence of quantum coherence affects non-adiabatic irreversible work in quadratic fermionic models.

In order to analyze the dynamical contribution of $Q_p^{[0,2]}$ to the average extracted work, we compare the r.h.s.~of Eq.~(\ref{eq:mean W}) with the average work obtained by applying the TPM scheme that considers the completely-dephased initial state $\Delta_1(\rho_0) = \sum_{n} \Pi_{t_1}^{(n)} \rho_0 \Pi_{t_1}^{(n)}$. The dephasing operator does not modify the initial average energy, but the absence of initial quantum coherence in the eigenbasis of $H_{t_1}$ unavoidably alters the final energy probabilities. Specifically, one has that
\begin{eqnarray}
    \displaystyle{ \mean{w[\rho_0]} -\mean{w[\Delta_1(\rho_0)]} = \frac{1}{L}\Tr{ \left[\rho_0 - \Delta_1(\rho_0)\right] H_{t_2}^{H} } }\nonumber\\ 
    \displaystyle{ =\int_{0}^{\pi}\frac{dk}{2\pi}\tanh \left(\frac{\beta \omega ^{[0]}_p}{2}\right)\omega^{[2]}_p  \left(Q^{[0,1]}_{p}Q_{p}^{[1,2]}-
    Q^{[0,2]}_{p}\right) }.\label{eq:DephasedAverageWorkDifference}
\end{eqnarray}
Hence, enhanced energy extraction can be obtained in a finite and connected region of parameters.
A significant advantage, with respect to what returned by the TPM scheme, is always achieved when $Q_p^{[0,1]}=0$ and $Q_p^{[0,2]}>0$. Remarkably, as shown in the next paragraph, these conditions are originated by the interplay between quantum coherences and quantum critical points.

\begin{figure}
    \centering
    \includegraphics[width=1\linewidth]{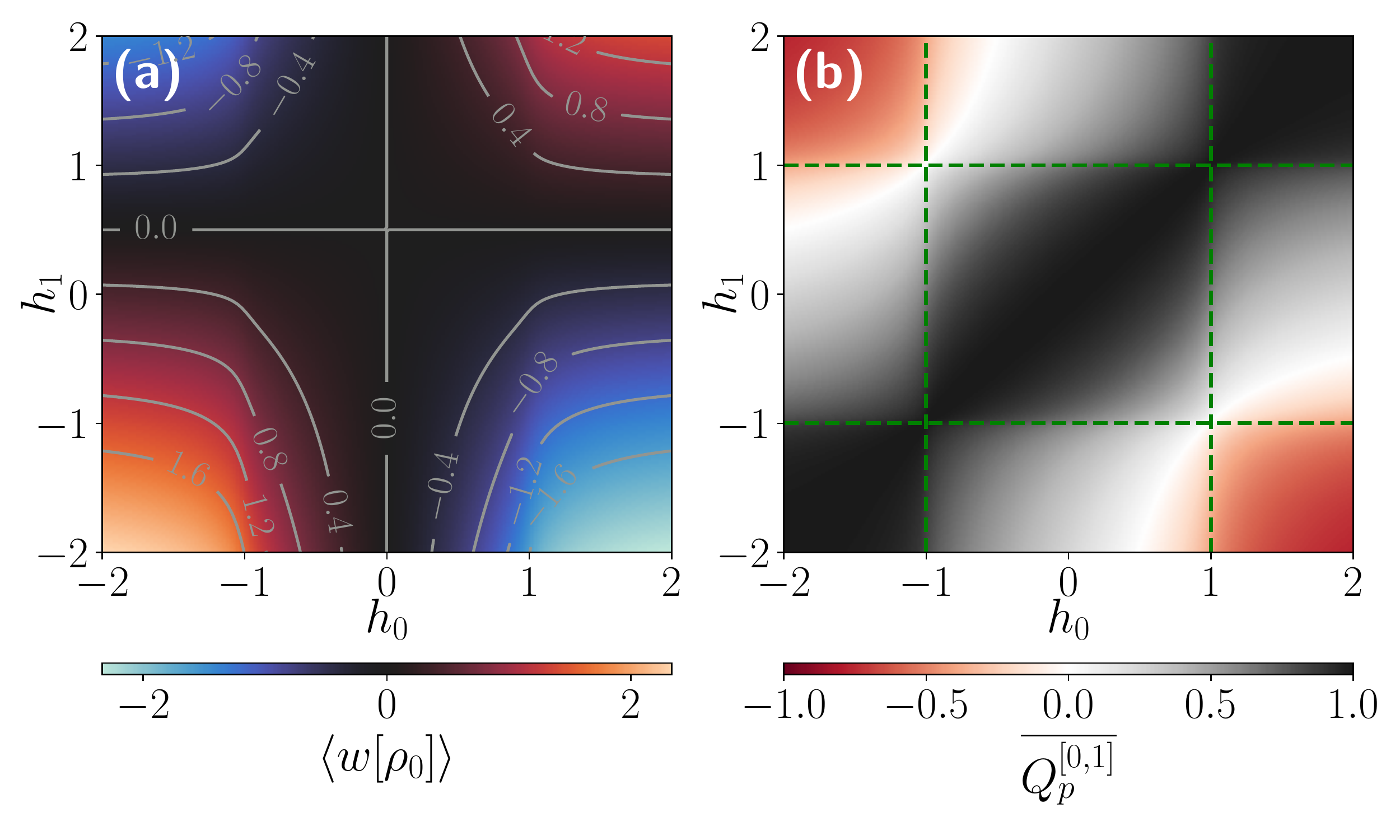}
    \caption{(a) Average work density originated by quenching the quantum Ising Hamiltonian from $h_1$ to $h_2=0.5$, for different values of $h_0$. The inverse temperature of the initial state $\rho_0 = \exp{-\beta H_{t_0}}/Z$ is taken equal to $\beta=15$. (b) Average overlaps between the eigenbases of $H_{t_1}$ and $H_{t_2}$.}
    \label{fig:CoherentWork}
\end{figure}
\begin{figure}
    \centering
    \includegraphics[width=.975\linewidth]{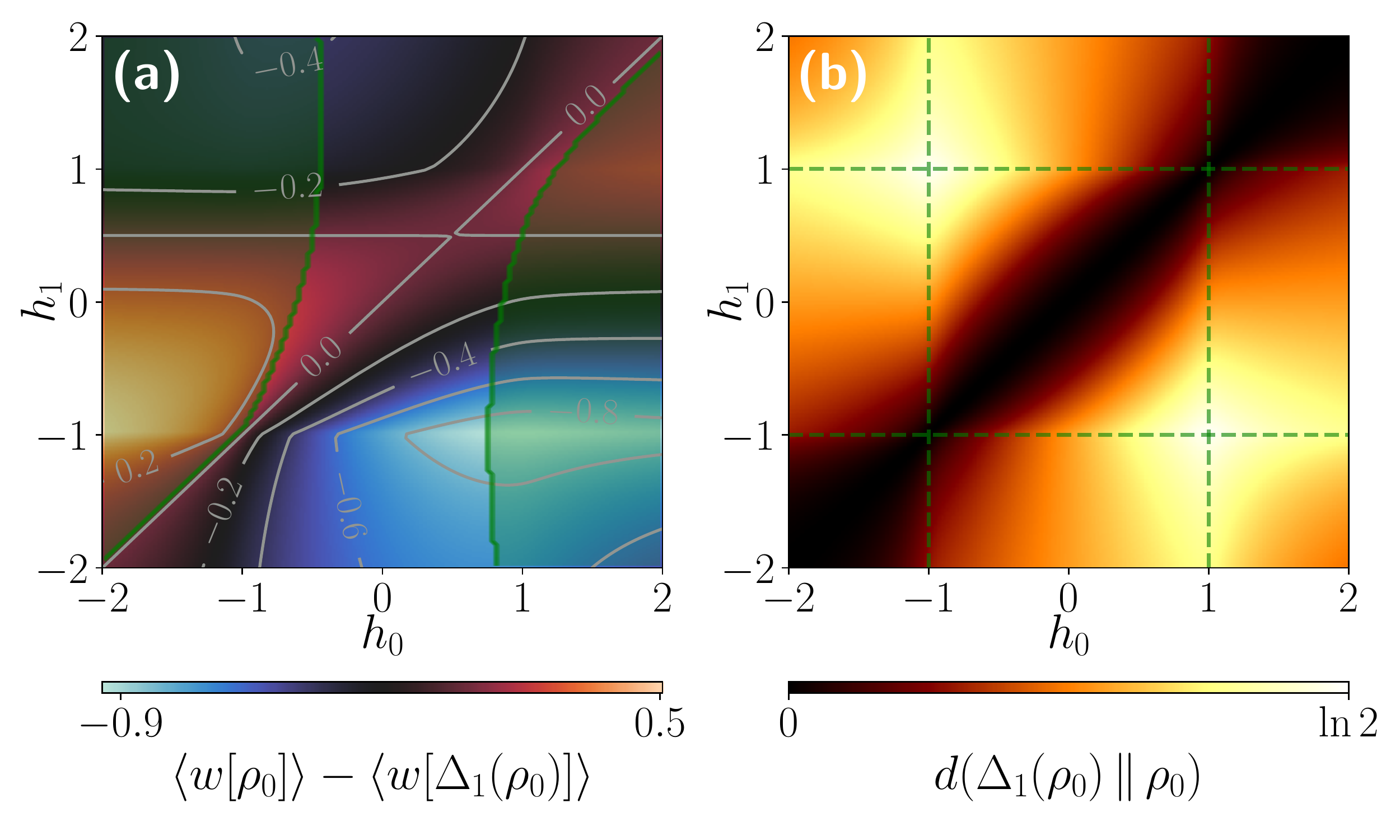}
    \caption{
    (a) Quantum correction to the average work density in the quantum Ising model, using the same Hamiltonian as in Fig.~\ref{fig:CoherentWork}(a), and enhancement of the amount of extracted work due to non-commutativity of the Hamiltonian operators at the different times of the work protocol. The shaded green areas mark the non-classical region of parameters where the fourth central moment of the KDQ distribution of work is negative. (b) Relative entropy of quantum coherences density, $D(\rho_0||\Delta_1(\rho_0))/L$, as a function of $h_0$ and $h_1$. The inverse temperature of the initial state is taken equal to $\beta=15$.
    }
    \label{fig:DeltaWork}
\end{figure}

\textbf{Enhanced energy extraction in quantum Ising model.}---To conclude the analysis, we show analytical and numerical results, concerning the enhancement of work extraction aided by non-classicality, in the concreate example of the quantum transverse field Ising model, which Hamiltonian reads as $H = -\sum_{j=1}^L \X_j\X_{j+1}+h\sum_{j=1}^L\Z_j$ where $\sigma^\alpha_j$ are the local Pauli matrices, with $\alpha=x,y,z$, $j=1,...,L$. The diagonalization of this model and its mapping to a fermionic model~\cite{Jordan1928berDP,mbeng2020quantum} are detailed in the SM. Following the protocol introduced above, we initialize the system in $\rho_0 = \exp{-\beta H_{t_0}}/Z$. $\rho_0$ is identified by the magnetic field $h=h_0$, while the Hamiltonian $H_{t_1}$ at the beginning of the work protocol is with $h=h_1$. 
Afterwards, we quench the external field from $h=h_1$ to $h=h_2$. Fig.~\ref{fig:CoherentWork}(a) shows the average work density exchanged during this driving process as a function of $h_0$ and $h_1$ with fixed $h_2=0.5$ and $\beta=15$. Energy extraction corresponds to negative values of the average work density. This has to be compared with Fig.~\ref{fig:CoherentWork}(b) where we plot $\overline{Q^{[0,1]}_p} = \int_0^{\pi}Q^{[0,1]}_{p}dp/\pi$ as an estimate of the average overlap between the eigenbases of $H_{t_0}$ and $H_{t_1}$. In agreement with our general analysis, the amount of energy extraction increases in the parameters region where $\overline{Q^{[0,1]}_p}$ is negative, i.e., the quantum coherences of $\rho_0$ in the basis of $H_{t_1}$ significantly change the work density. Interestingly, the optimal region for energy extraction corresponds to $h_0>1$ in the ferromagnetic phase and $h_1<-1$ in the antiferromagnetic phase or vice versa. In Fig.~\ref{fig:DeltaWork}(a) we plot the right-hand-side of Eq.~\eqref{eq:DephasedAverageWorkDifference} as a function of $h_0$ and $h_1$, with $h_2=0.5$ and $\beta=15$. It represents the quantum correction to the average work density, provided by an extra contribution (than the TPM scheme) due to non-commutativity. In this regard, in Fig.~\ref{fig:DeltaWork}(a) we also highlight a region (green shaded area) where the sign of the fourth central moment of the work distribution is negative. This signals the presence of a non-classical region corresponding to the fact that the KDQ distribution of work has negative values. Accordingly, in such non-classical region, $\mean{w[\rho_0]}<0$ witnesses an enhancement of energy extraction that is boosted by quantum coherence of the initial state $\rho_0$, expressed in the basis of $H_{t_1}$.

For quadratic fermionic models, we have already argued that $Q_p^{[0,1]}=0$ and $Q_p^{[0,2]}>0$ always allow for energy extraction enhancement. In Fig.~\ref{fig:CoherentWork}(b) [details are in the SM], we specialise our analysis to the quantum Ising model where the maximum enhancement in energy extraction Eq.~\eqref{eq:DephasedAverageWorkDifference} is exactly satisfied by choosing $h_0 = \pm 1$  and $h_1 = \mp 1$, i.e., in parameter regions where the initial state and the initial Hamiltonian sit at the two opposite quantum critical points of the model. Remarkably, these points also correspond to the maximum of the {\it relative entropy of quantum coherences}~\cite{BaumgratzPRL2014,StreltsovRevModPhys2017} that is defined as
\begin{align}
    D(\rho_0||\Delta_1(\rho_0)) = 
    %
    %
    \Tr{ \rho_0 \Big( \ln\rho_0 - \ln\Delta_1(\rho_0) \Big) }.
\end{align} 
Its density $d(\rho_0||\Delta_1(\rho_0))=D(\rho_0||\Delta_1(\rho_0))/L$
is thus plotted in Fig.~\ref{fig:DeltaWork}(b) as a function of $h_0$ and $h_1$. Consequently, the correspondence between the maximum value of $D(\rho_0||\Delta_1(\rho_0))$ and the enhancement of the work extraction benchmarks that this advantage originates from the non-commutativity of the initial state $\rho_0$ with the Hamiltonian $H_{t_1}$ at the beginning of the work protocol. At the critical points of the quantum Ising model, it becomes a quite useful thermodynamic resource.

\begin{figure}
    \centering
    \includegraphics[width=\linewidth]{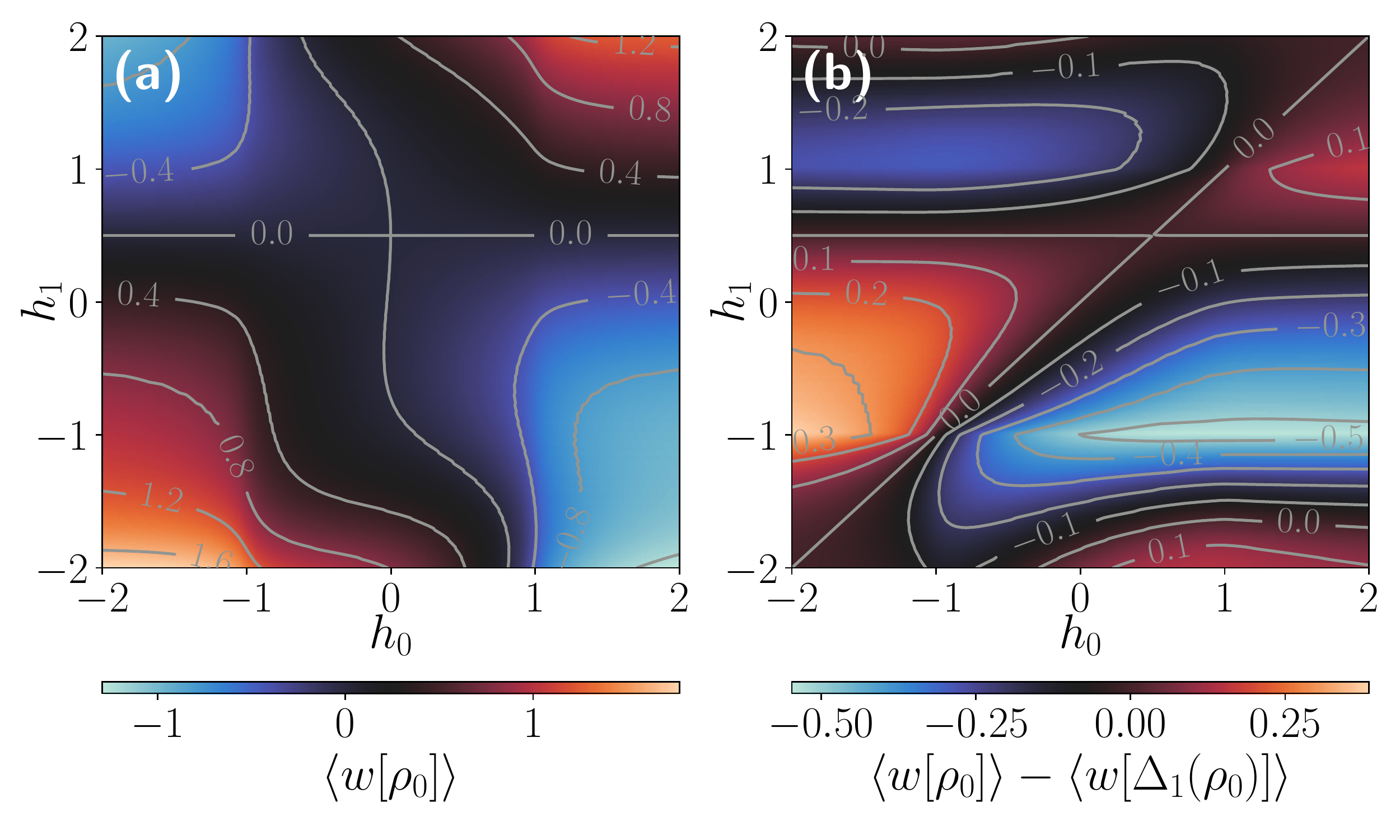}
    \caption{Average work density (a) and enhancement of extracted work (b), obtained by changing the Hamiltonian with a linear ramp from $h_1$ to $h_2=0.5$. The finite velocity of the driving is fixed to $\delta=4$, for different values of $h_0$. Again, the inverse temperature of the initial state is $\beta=15$.
    }
    \label{fig:LZSM_Quench}
\end{figure}

The above results are obtained by assuming a sudden change of the initial Hamiltonian with a quench dynamics. However, the validity of such results can be also confirmed by changing the Hamiltonian of the work protocol via a sufficiently fast, but time-finite, ramp drive. Specifically, we have taken the latter equal to $h(t)=h_1 + \delta (t-t_1)$, with $\delta=(h_2-h_1)/(t_2-t_1)$. In this case the system evolution is still exactly solvable since it can be decomposed in the dynamics of $L$ independent Landau-Zener-St{\"u}ckelberg-Majorana models~\cite{Vitanov1996PRA,DziarmagaAdvancesinPhysics2010} one for each Fourier mode.
 The resulting average work density [Eq.~\eqref{eq:mean W}] and enhanced extracted work [Eq.~\eqref{eq:DephasedAverageWorkDifference}], attained for $\delta=4$ are plotted in Fig.~\ref{fig:LZSM_Quench}. For completeness, further exact results derived for driving functions at finite velocities are in the SM. 

\textbf{Conclusions.}---We have analyzed a fundamental aspect of quantum thermodynamics, especially when applied to many-body systems, namely the understanding of how non-commutativity between quantum states and the Hamiltonian of the system, evaluated over two times, enters work fluctuations. To answer this question, we make use of the Kirkwood-Dirac
quasiprobability approach introduced for two-times quantum correlators, by obtaining analytical results for a generic quadratic fermionic model, among which the transverse field Ising model can be easily cast. Beyond determining a correction to the work statistics strictly depending on non-commutativity, the main outcome of our analysis is to show a clear connection between the following concepts: {\it i)} non-classical signatures in the KDQ distribution of work; {\it ii)} enhanced work extraction under non-adiabatic coherent driving and non-commutativity of $\rho_0$ and $H_t$ during the work protocol; {\it iii)} critical behaviours in quantum fermionic systems across different matter phases. We believe that our study could open a novel research line for determining non-classical fluctuations of thermodynamic quantities in a generic quantum many-body system thanks to the use of KDQ distributions. Moreover, we also propose the application of a quasi-probability approach (e.g., the KDQ) to shed light with a novel perspective 
at the quantum nature of the phase transitions occurring in spin and fermionic many-body systems.
Finally, in line with Refs.~\cite{CampisiNatComm2016,SolfanelliArXiv2022}, our results about the enhancement of work extraction would worth further investigation for the design of quantum heat engines~\cite{Campisi_2015,Solfanelli2021PRXQ,piccitto2022ising,CangemiArXiv2023} and quantum batteries~\cite{CampaioliShortReview,RossiniPRL2020} powered by genuinely quantum features and showing a clear quantum advantage.

\begin{acknowledgments}
This work was supported by the European Commission under GA n.~101070546--MUQUABIS (S.G.), and the PNRR MUR project PE0000023-NQSTI (M.C. and S.G.). 

\textit{Note added.}---While completing this work, the preprint~\cite{Farncica2023ArXiv} appeared, dealing with topics related to the ones discussed by us.
\end{acknowledgments}

\clearpage
\setcounter{equation}{0}
\renewcommand{\theequation}{S.\arabic{equation}}
\setcounter{figure}{0}
\renewcommand{\thefigure}{S.\arabic{figure}}

\begin{widetext}
    
\begin{center}
     \textbf{\Large Supplemental Material}
\end{center}
\appendix

\section*{From quantum Ising model to quadratic fermionic Hamiltonians}

The Hamiltonian operator of the transverse field quantum Ising model reads
\begin{equation}\label{SM_eq_H_Ising}
 H_t = -J \sum_{j=1}^L \X_j\X_{j+1}+h(t)\sum_{j=1}^L\Z_j,
\end{equation}
where $\sigma^\alpha_j$ are the local Pauli matrices, with $\alpha=x,y,z$ and $j=1,...,L$. It thus holds that $\comm*{\sigma^\alpha_i}{\sigma^\beta_j} = 2 i\delta_{ij} \epsilon^{\alpha\beta\gamma}\sigma_j^\gamma$, where $\delta$ denotes the Kronecker delta and $\epsilon$ is the Levi-Civita symbol. We set $J=1$ in order to fix the energy scale of the system. At zero temperature, the model has a ferro-/para-magnetic phase transition for $h=\pm1$. As it is customary, we transform the Hamiltonian (\ref{SM_eq_H_Ising}) by means of the Jordan-Wigner transformation \cite{Jordan1928berDP,mbeng2020quantum}
\begin{equation}
     \X_{k} = \prod_{j=1}^{k-1}(1-2  n_{j}) ( c^{\dag}_{k}+ c_{k} ), \quad \sigma^{y}_{k} = i\prod_{j=1}^{k-1}(1-2  n_{j}) ( c^{\dag}_{k} -  c_{k}), \quad
 \sigma^{z}_{k} = 1 - 2  n_{k} \,,
\end{equation}
where $\{ c_{i}, c^{\dag}_{j}\} = \delta_{ij}$ and $n_j = c^\dagger_jc_j$. After the Jordan-Wigner transformation, the Hamiltonian takes the following form:
\begin{equation}
H_t = -\sum_{j=1}^{L} (c^{\dag}_{j} c_{j+1} + c_j^\dag c_{j+1}^\dag +\mathrm{h.c.})-h(t)\sum_{j=1}^L(2 c^\dag_jc_j-1).
\end{equation}
Since the model is translationally invariant, the Hamiltonian $H_t$ can be diagonalized by means of the discrete Fourier transform
\begin{equation}
    c_j = \frac{e^{-i\pi/4}}{\sqrt{L}}\sum_{p} e^{ipj} \tilde{c}_p \,, \quad \quad \tilde{c}_p = \frac{e^{i\pi/4}}{\sqrt{L}}\sum_{j=1}^L e^{-ipj} c_j
\end{equation}
with $p=2\pi m/L$ and $m =-L/2+1,\dots,L/2$. Moreover, the invariance under the inversion symmetry $p\to-p$ allows us to restrict the computations to positive momenta by defining $\Psi_p = ( \tilde{c}_p, \tilde{c}^\dagger_{-p})^T$. Therefore,
\begin{equation}
    H_t = \sum_{p > 0} \Psi_p^\dagger \mathbb{H}_p(t) \Psi_p
\end{equation}
where
\begin{equation}
    \mathbb{H}_p(t) = \begin{pmatrix*}[c]
    -2\cos p + 2h(t) & -2\sin p \\
     -2\sin p  & 2\cos p - 2h(t)
    \end{pmatrix*}.
\end{equation}
The $R_y(\phi_p(t)) = {\rm exp}(i\phi_{p}\sigma^y/2)$ rotation applied to the new fermions $\Gamma_p(t) = (\gamma_p, \gamma^\dagger_{-p})^T$ diagonalizes the problem. Formally, one has
\begin{equation}
\Psi_p = \exp(i\frac{\phi_p(t)}{2} \sigma^y) \Gamma_p(t)\equiv R_{y}(\phi_p(t)) \Gamma_p(t),
\end{equation} 
where the eigenvectors of the rotation are given by
\begin{equation}\label{eq:eigenvectors}
v_{+,p} = \begin{pmatrix*}[l]
    \cos\phi_p(t)/2 && \sin\phi_p(t)/2
\end{pmatrix*}^T \quad \text{and} \quad
v_{-,p} = \begin{pmatrix*}[l]
    -\sin\phi_p(t)/2 &&\cos\phi_p(t)/2
\end{pmatrix*}^T \,. 
\end{equation}
The rotation angles $\phi_p$ are implicitly defined by the conditions $\cos\phi_p(t)=2(h(t)-\cos{p})/\omega_p(t)$ and $\sin\phi_p(t) = 2\sin{p}/\omega_p(t)$; note that $\phi_{p}(t) = -\phi_{-p}(t)$. 
The Hamiltonian, written in terms of the new fermionic operators, then reads
\begin{equation}
    H_t = \sum_{p>0 } \Gamma_p^\dagger(t) \mathbb{D}_p(t) \Gamma_p(t) = \sum_{p>0} \Psi_p^\dagger R^\dagger_y(\phi_p(t))\omega_p(t)\sigma^z R_y(\phi_p(t))\Psi_p
    \label{eq:Bogoliubov_diagonalized}
\end{equation}
where $\mathbb{D}_p(t) = \omega_p(t) \sigma^z $ and the energies of each mode are given by $\omega_p(t) = 2\sqrt{(\cos p-h(t))^2+\sin^2 p}$, with $\omega_{p}(t) = \omega_{-p}(t)$.

\section*{Derivation of the average work}

In this section we show an alternative derivation of equation \eqref{eq:mean W} in the main text by directly computing
\begin{equation}
    %
    \mean{W_{[t_1,t_2]}} = \Tr{U\rho_0U^\dagger H_{t_2}}-\Tr{\rho_0 H_{t_1}}.
\end{equation}
First of all, we consider the Hamiltonian $H_t$ in diagonalized form, i.e., \begin{equation}\label{app:Hamiltonian_diagonalized}
    H_{t_j} =\sum_{p>0} \Psi^\dagger_p \mathbb{H}^{[j]}_p\Psi_p = \sum_{p>0} \omega_p^{[j]}\left({\gamma_p^{[j]}}^\dagger\gamma^{[j]}_p-\gamma^{[j]}_{-p}{\gamma^{[j]}_{-p}}^\dagger\right).
\end{equation}
Then, from Eq.~(\ref{app:Hamiltonian_diagonalized}), the initial state can be written as
\begin{equation}
\rho_0 = \frac{e^{-\beta H_{t_0}}}{Z^{[0]}} = \prod\limits_{p>0} \frac{\exp{-\beta\omega_p^{[0]}\left(n^{[0]}_p + n^{[0]}_{-p} -1\right)}}{Z^{[0]}_p} \equiv \prod\limits_{p>0} \rho_p^{[0]},
\end{equation}
where we have defined the number operator in the $p$-momentum subspace $n_{\pm p}^{[0]} = {\gamma_{\pm p}^{[0]}}^\dagger\gamma^{[0]}_{\pm p}$, and $Z^{[0]} = \Tr{e^{-\beta H^{[0]}}}=\prod\limits_{p>0} Z^{[0]}_p = \prod\limits_{p>0} 2\left(1+\cosh(\beta\omega^{[0]}_p)\right)$. In order to compute the average work, we have to consider\begin{align}
   &\Tr{\rho_0U^\dagger H^{[2]} U} = \prod_{q>0} \Tr{\rho^{[0]}_q\sum_{p>0}{H_p^{[2]}}^\mathrm{H} } = \sum_{p>0} \Tr{\rho_p^{[0]} \Psi^\dagger_p \mathbb{U}_{p,t_{1}:t_{2}}^\dagger R^\dagger_y(\phi_p^{[2]}) \omega_p^{[2]}\sigma^z R_y(\phi_p^{[2]})\mathbb{U}_{p,t_{1}:t_{2}}\Psi_p } \\
   & =\sum_{p>0} \Tr{\rho_p^{[0]} {\Gamma^{[0]}_p}^\dagger R^\dagger_y(\phi_p^{[0]})\mathbb{U}_{p,t_{1}:t_{2}}^\dagger R^\dagger_y(\phi_p^{[2]}) \omega_p^{[2]}\sigma^z R_y(\phi_p^{[2]})\mathbb{U}_{p,t_{1}:t_{2}}R_y(\phi_p^{[0]})\Gamma^{[0]}_p } \equiv \sum_{p>0}\Tr{\rho^{[0]}_p{\Gamma_p^{[0]}}^\dagger \mathbb{M} \Gamma_p^{[0]}} \notag
\end{align}
with $\Psi_p = R_y(\phi_p^{[0]})\Gamma_p^{[0]}$ and $\mathbb{M} = R^\dagger_y(\phi_p^{[0]})\mathbb{U}_{p,t_{1}:t_{2}}^\dagger R^\dagger_y(\phi_p^{[2]}) \omega_p^{[2]}\sigma^z R_y(\phi_p^{[2]})\mathbb{U}_{p,t_{1}:t_{2}}R_y(\phi_p^{[0]})$. Then, by computing the trace over the Fock space $\{\ket{\emptyset},\gamma^\dagger_p\ket{\emptyset},\gamma^\dagger_{-p}\ket{\emptyset},\gamma^\dagger_p\gamma^\dagger_{-p}\ket{\emptyset}\}$, we can find that
\begin{align}
    \Tr{\rho^{[0]}_p{\Gamma_p^{[0]}}^\dagger \mathbb{M} \Gamma_p^{[0]}} &= \mathbb{M}_{11} \Tr{\rho^{[0]}_p{\gamma^{[0]}_p}^\dagger\gamma^{[0]}_p}+\mathbb{M}_{12} \Tr{\rho^{[0]}_p{\gamma^{[0]}_p}^\dagger {\gamma^{[0]}_{-p}}^\dagger}+\mathbb{M}_{21} \Tr{\rho^{[0]}_p\gamma^{[0]}_p\gamma^{[0]}_{-p}}+\mathbb{M}_{22} \Tr{\rho_p^{[0]}\gamma^{[0]}_{-p} {\gamma^{[0]}_{-p}}^\dagger }\notag\\
    &= \mathbb{M}_{11}\Tr{\rho_p^{[0]}n^{[0]}_p}+ \mathbb{M}_{22}\Tr{\rho^{[0]}_p\left(1-n_{-p}^{[0]}\right)} = \mathbb{M}_{11} \frac{1+e^{-\beta\omega_p^{[0]}}}{Z_p^{[0]}} + \mathbb{M}_{22} \frac{e^{\beta\omega_p^{[0]}} +1 }{Z_p^{[0]}}\,.
\end{align}
From the cyclic property of the trace, it is easy to check that $\tr{\mathbb{M}}=0$, and therefore $\mathbb{M}_{11} = - \mathbb{M}_{22}$ whereby
\begin{equation}
    \Tr{\rho^{[0]}_p{\Gamma_p^{[0]}}^\dagger \mathbb{M} \Gamma_p^{[0]}} = \mathbb{M}_{22} \frac{e^{\beta\omega_p^{[0]}}-e^{-\beta\omega_p^{[0]}}}{Z^{[0]}_p} = \frac{\mathbb{M}_{22}}{2} \tanh{\frac{\beta\omega_p^{[0]}}{2}}\,.
\end{equation}
If we define $\hat{e}_+ = (1\ \ 0)^T$ and $\hat{e}_-=(0\ \ 1)^T$, $\mathbb{M}_{22}$ can be expressed as
\begin{equation}
    \mathbb{M}_{22} =  \omega_p^{[2]} \left[\hat{e}_-^T R^\dagger_y(\phi_p^{[0]})\mathbb{U}_{p}^\dagger(t) R^\dagger_y(\phi_p^{[2]})(\hat{e}_+\cdot\hat{e}_+^T-\hat{e}_-\cdot\hat{e}_-^T)R_y(\phi_p^{[2]})\mathbb{U}_{p}(t)R_y(\phi_p^{[0]}) \hat{e}_-\right].
\end{equation}
From the explicit expression of the eigenvectors in equations \eqref{eq:eigenvectors},
$R(\phi_p^{[j]}) = \begin{pmatrix*}[c]
        v_{+,p}^{[j]} && \vline && v_{-,p}^{[j]}
    \end{pmatrix*}$, so that $R(\phi_p^{[j]}) \hat{e}_i = v^{[j]}_{i,p}$ with $i=\pm$. Therefore,
\begin{equation}
\mathbb{M}_{22} = \omega_p^{[2]}\left[\left|{v_{+,p}^{[2]}}^{\dagger}\mathbb{U}_{p,t_{1}:t_{2}}v^{[0]}_{-,p}\right|^2 - \left|{v_{-,p}^{[2]}}^{\dagger}\mathbb{U}_{p,t_{1}:t_{2}}v^{[0]}_{-,p}\right|^2 \right] = -\omega_p^{[2]}\left[P_p^{[0,2]} - (1-P_p^{[0,2]})  \right] = -\omega_p^{[2]}\left(2P_p^{[0,2]} - 1  \right),
\end{equation}
where $P_p^{[0,2]}$ denotes the probability of not transitioning between the instantaneous eigenstates of $H^{[2]}_p$.
It is worth pointing out that the transition matrix with elements $\mathcal{P}_{i,j}^{[0,2]} = \abs{{v_{i,p}^{[2]}}^\dagger \mathbb{U}_{p,t_{1}:t_{2}} v_{j,p}^{[0]} }^2$ is bistochastic, which implies that $\sum_i \mathcal{P}_{i,j}^{[0,2]} = \sum_j\mathcal{P}_{i,j}^{[0,2]} = 1$. As a result, we can define
\begin{equation}    \abs{{v_{+,p}^{[2]}}^{\dagger}\mathbb{U}_{p,t_{1}:t_{2}}v^{[0]}_{+,p}}^2=\abs{{v_{-,p}^{[2]}}^{\dagger}\mathbb{U}_{p,t_{1}:t_{2}}v^{[0]}_{-,p}}^2=P^{[0,2]}_p \quad \text{and} \quad \abs{{v_{+,p}^{[2]}}^{\dagger}\mathbb{U}_{p,t_{1}:t_{2}}v^{[0]}_{-,p}}^2=\abs{{v_{-,p}^{[2]}}^{\dagger}\mathbb{U}_{p,t_{1}:t_{2}}v^{[0]}_{+,p}}^2= 1-P^{[0,2]}_p.
\end{equation}
Moreover, $\Tr{\rho_0 H^{[1]}}$ is obtained setting $\mathbb{U}_{p,t_{1}:t_{2}}=\mathbb{I}_p$. Hence, substituting $2\to1$ concludes our derivation of the average work that leads to
\begin{equation}
    \mean{W_{[t_1,t_2]}}= \frac{L}{2\pi} \int_0^{\pi}dp\tanh\left(\frac{\beta\omega_p^{[0]}}{2}\right)\left[\omega_p^{[1]} \left(2P_p^{[0,1]}-1\right)-\omega_p^{[2]}\left(2P_p^{[0,2]}-1\right)\right].
\end{equation}
The formula above reduces to the expression for the average work shown in the main text upon introducing the parameters $Q_p^{[i,j]} = 2P_p^{[i,j]}-1$. More details on the properties of such parameters as functions of the Fourier modes and for different values of the chemical potential $h$ are provided in the next section. 

\section*{Transition probabilities and Bogoliubov angles}
\begin{figure}
    \centering
    \includegraphics[width=\linewidth]{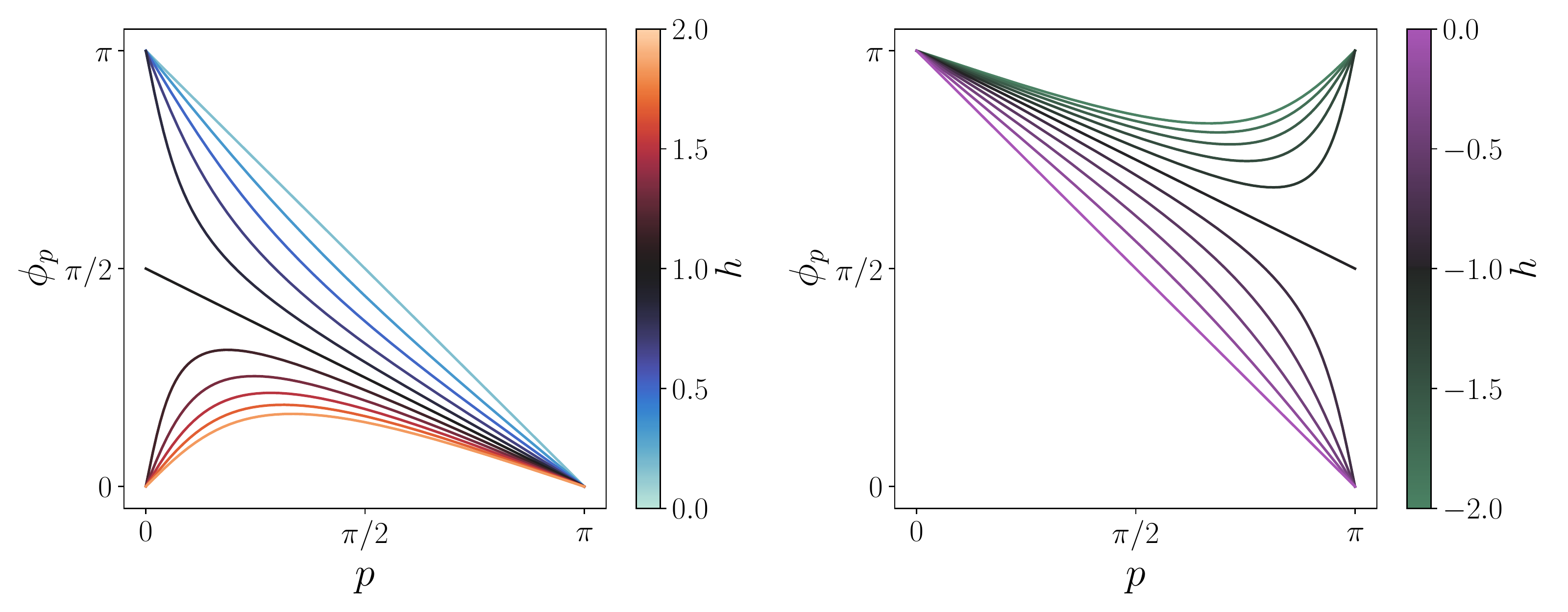}
    \caption{Rotation angles for different values of $h$ as a function of $p$; left panel, ferromagnetic phase; right panel, paramagnetic phase.}
    \label{fig:rotation angles}
\end{figure}
We here provide a detailed analysis of the properties of both the transition probabilities $P_p^{[i,j]}$ and the parameters $Q_p^{[i,j]} = (2P_p^{[i,j]}-1)$ entering the expression for the average work in the main text. Let us start from $Q_p^{[0,1]}$; this quantity describes the overlap between the eigenbasis of the initial Hamiltonian $H_{t_1}$, with chemical potential $h_1$, and the eigenbasis of the initial state $\rho_0$. The latter is chosen of the form $\rho_0 = e^{-\beta H_{t_0}}/Z$, where $H_{t_0}$ is a quadratic fermionic Hamiltonian with chemical potential $h_0$. Accordingly, $Q_p^{[0,1]}$ can be written in terms of the Bogoliubov angles $\phi_p^{[0]},\phi_p^{[1]}$ associated to the rotations that diagonalize the Hamiltonians $H_{t_{0}}$ and $H_{t_1}$, i.e.,
\begin{align}
    Q_p^{[0,1]} = \cos(\phi_p^{[0]}-\phi_p^{[1]}).
\end{align}
It thus follows that the properties of $Q_p^{[0,1]}$ derive directly from the properties of the Bogoliubov angles. Figure \ref{fig:rotation angles} shows the Bogoliubov angles as a function of the Fourier modes $p$ in the different phases of the model, i.e., for different values of the control parameter $h$. One can note that the different phases of the model can be identified by behavior of the Bogoliubov angles at $p = 0,\pi$. In fact, for $p\to 0$ we have that
\begin{align}
    \lim_{p\to 0}\phi_p = \begin{cases}
        0 &\mathrm{if}\quad h>1\\
        \pi/2 &\mathrm{if}\quad h=1\\
       \pi       &\mathrm{if}\quad h<1 \,,\\
    \end{cases}
\end{align}
with a jump discontinuity at the critical point $h = 1$. Similarly, for $p\to\pi $ we find
\begin{align}
    \lim_{p\to \pi}\phi_p =
\begin{cases}
        0 &\mathrm{if}\quad h>-1\\
        \pi/2 &\mathrm{if}\quad h=-1\\
       \pi       &\mathrm{if}\quad h<-1 \,,\\
    \end{cases}
\end{align}
with another discontinuity in correspondence of the second critical point $h = -1$. Moreover, notice that for $h = 0,\pm 1$ the Bogoliubov angle $\phi_p$ takes the simple form 
\begin{align}
   \phi_p = \begin{cases}
         \pi/2-p/2     &\mathrm{if}\quad h = 1\\
         \pi-p&\mathrm{if}\quad h=0\\
        -p/2&\mathrm{if}\quad h=-1 \,.
    \end{cases}
\end{align}
These properties of the Bogoliubov angles translate into properties of $Q_p^{[0,1]}$, which are estimated in the main text through their average over the Fourier modes: $\overline{Q_p^{[i,j]}} = \frac{1}{\pi}\int_0^\pi Q_p^{[i,j]} dp$.

\section*{Symmetries of the average work}\label{SMsec:Simmetries}

In this section we provide a proof that the average work $\mean{W_{[t_1,t_2]}}$ (from here on simply denoted with $\mean{W(h)}$ unless specified) is an {\it odd function} of $h_0$, i.e.,  
\begin{equation}
\mean{W(h_0)} = -\mean{W(-h_0)}.
\end{equation}
For the sake of presentation, we are going to write explicitly the dependence on $h(t)$, namely $h_j = h(t_j)$. 

Let us thus start from a summary of properties of both the single particle spectrum and the Bogoliubov angles, which are necessary to carry out the proof. In particular, directly from their definitions in the main text, we find that
\begin{align}
    \omega_{p}(h) = \omega_{-p}(h) = \omega_{\pi-p}(-h),
\end{align}
while for the Bogoliubov angles it holds that
\begin{align}
    \phi_p(h) = -\phi_{-p}(h) = \pi-\phi_{\pi-p}(-h).
\end{align}
With the above properties in mind, we consider the average work that can be expressed as 
\begin{align}
    \mean{W(h_0)}= \frac{L}{2\pi} \int_0^{\pi}dp\tanh\left(\frac{\beta\omega_p(h_0)}{2}\right)\left[\omega_p(h_1) \cos\left(\phi_p(h_0)-\phi_p(h_1)\right)-\omega_p(h_2)\cos\left(\phi_p(h_0)-\phi_p(h_2)\right)\right],
\end{align}
where we have used the fact that $Q_{p}^{[i,j]} = \cos\left(\phi_p(h_i)-\phi_p(h_j)\right)$ for quench dynamics. To proceed with our proof, it is convenient to translate the integration variable by $\pi/2$. Thus, we introduce the change of variables $k = p+\pi/2$, so that
\begin{align}
    \mean{W(h_0)}= \frac{L}{2\pi} \int_{\frac{\pi}{2}}^{\frac{3\pi}{2}}dp\tanh\left(\frac{\beta\omega_{k-\frac{\pi}{2}}(h_0)}{2}\right)[&\omega_{k-\frac{\pi}{2}}(h_1) \cos\left(\phi_{k-\frac{\pi}{2}}(h_0)-\phi_{k-\frac{\pi}{2}}(h_1)\right)\notag\\
    -&\omega_{k-\frac{\pi}{2}}(h_2)\cos\left(\phi_{k-\frac{\pi}{2}}(h_0)-\phi_{k-\frac{\pi}{2}}(h_2)\right)].
\end{align}
Then, exploiting that $\omega_{k-\frac{\pi}{2}}(h_0) = \omega_{\frac{\pi}{2}-k}(-h_0) = \omega_{k-\frac{\pi}{2}}(-h_0)$ and $\phi_{k-\frac{\pi}{2}}(h_0) = \pi-\phi_{\frac{\pi}{2}-k}(-h_0) = \pi+\phi_{k-\frac{\pi}{2}}(-h_0)$, we determine that
\begin{align}
    \mean{W(h_0)} = \frac{L}{2\pi} \int_{\frac{\pi}{2}}^{\frac{3\pi}{2}}dp\tanh\left(\frac{\beta\omega_{k-\frac{\pi}{2}}(-h_0)}{2}\right)[&\omega_{k-\frac{\pi}{2}}(h_1) \cos\left(\pi+\phi_{k-\frac{\pi}{2}}(-h_0)-\phi_{k-\frac{\pi}{2}}(h_1)\right)
    \notag\\
    -&\omega_{k-\frac{\pi}{2}}(h_2)\cos\left(\pi+\phi_{k-\frac{\pi}{2}}(-h_0)-\phi_{k-\frac{\pi}{2}}(h_2)\right)].
\end{align}
Finally, using the trigonometric identity $\cos(\pi+x) = -\cos(x)$ and translating back the integration variable to $p = k-\pi/2$, we obtain
\begin{align}
    \mean{W(h_0)} &= -\frac{L}{2\pi} \int_{0}^{\pi}dp\tanh\left(\frac{\beta\omega_{p}(-h_0)}{2}\right)[\omega_{p}(h_1) \cos\left(\phi_{p}(-h_0)-\phi_{p}(h_1)\right)
    -\omega_{p}(h_2)\cos\left(\phi_{p}(-h_0)-\phi_{p}(h_2)\right)]\notag \\ 
    &= -\mean{W(-h_0)}.
\end{align}
This concludes our proof.

\section*{Average work with initial dephased quantum state}

In this section we aim to extend the above discussion to the case of a dephased quantum state $\Delta_{1}(\rho_0)\equiv \sum_\alpha\Pi^{[1]}_\alpha \rho_0 \Pi^{[1]}_\alpha$, where $\Pi_\alpha^{[1]}$ are projectors in the $H^{[1]}$ basis. In this case, we find that the corresponding average work, identified by $\mean{W[\Delta_1(\rho_0)]}$, equals to
\begin{equation}
    \mean{W[\Delta_1(\rho_0)]} = \Tr{\Delta_1(\rho_0)U^\dagger H_{t_2} U}- \Tr{\Delta_1(\rho_0)H_{t_1}} =\Tr{\Delta_1(\rho_0)U^\dagger H_{t_2} U}- \Tr{\rho_0H_{t_1}}.
\end{equation}
In order to find an explicit expression for the average work originated by a work protocol starting from $\Delta_1(\rho_0)$, we need to project the state $\rho_0$ onto the basis that decompose the initial Hamiltonian $H_{t_{1}}$. For this purpose, we consider the state with
\begin{equation}
    Z^{[0]}_p\rho^{[0]}_p =  e^{\beta\omega_p^{[0]}}\left\{\left[1+n_p^{[0]}\left(e^{-\beta\omega_p^{[0]}}-1\right)\right]\left[1+n_{-p}^{[0]}\left(e^{-\beta\omega_p^{[0]}}-1\right)\right]\right\}.
\end{equation}
Then, we have to rewrite such state in terms of the fermionic operators that diagonalize $H_{t_1}$. The latter are related to the fermionic operators diagonalizing $\rho_0$ through the condition
\begin{equation}
        \gamma_p^{[0]} = C_p \gamma_p^{[1]}+S_p{\gamma_{-p}^{[1]}}^\dagger \,,
\end{equation}
where 
\begin{equation}
    C_p = \cos(\frac{\phi^{[1]}_p-\phi^{[0]}_p}{2}) \quad \text{and} \quad S_p = \sin(\frac{\phi^{[1]}_p-\phi^{[0]}_p}{2}).
\end{equation}
Therefore, the number of $0$-fermions in the $p$-mode can be expressed in terms of $\gamma^{[1]}_p$, i.e.,
\begin{equation}
    n^{[0]}_p = \left[ C_p^2 n^{[1]}_{p} + S^2_p(1-n^{[1]}_{-p})\right] + C_pS_p\left[{\eta^{[1]}_p}^\dagger+{\eta^{[1]}_p}\right],
\end{equation}
where we have introduced the bosonic operator $\eta_p = \gamma_p\gamma_{-p}$. Thus, using the relations above, we can write $\rho_0$ in the basis of $H_{t_1}$:
\begin{align}
    Z^{[0]}_{p}\rho_p^{[0]} = e^{\beta\omega_p^{[0]}}&\biggl\{\biggl[C^2_p\left(1+n_p^{[1]}\left(e^{-\beta\omega_p^{[0]}}-1\right)\right)+S^2_p\left(1+(1-n_{-p}^{[1]})\left(e^{-\beta\omega_p^{[0]}}-1\right)\right)+\notag\\
    &+C_pS_p\left({\eta^{[1]}_p}^\dagger+{\eta^{[1]}_p}\right)(e^{-\beta\omega_p^{[0]}}-1) \biggr]\left[\left(p\to -p\right)\right]\biggr\};
\end{align}
notice that $S_{-p}=-S_p$. In the following, in order to extract only the diagonal contribution, we are going to discard the terms that are not diagonal in $n^{[1]}_p, n^{[1]}_{-p}$. Accordingly, by defining $\rho_p^{[1|0]}=\sum_{\alpha} \Pi^{[1]}_\alpha \rho_p^{[0]} \Pi^{[1]}_\alpha$, we find that
\begin{align}
     &Z^{[0]}_{p}\rho_p^{[1|0]} = \biggl[C_p^4 e^{-\beta\omega_p^{[0]}\left(n_p^{[1]}+n_{-p}^{[1]}-1\right)} + S_p^4  e^{\beta\omega_p^{[0]}\left(n_p^{[1]}+n_{-p}^{[1]}-1\right)} +2S_p^2 C_p^2-(Z^{[0]}_p-4)S^2_pC^2_p\left({\eta^{[1]}_p}^\dagger+{\eta^{[1]}_p}\right)\left({\eta^{[1]}_{-p}}^\dagger+{\eta^{[1]}_{-p}}\right)\biggr] \notag\\
      &=\left[C_p^4 e^{-\beta\omega_p^{[0]}\left(n_p^{[1]}+n_{-p}^{[1]}-1\right)} + S_p^4  e^{\beta\omega_p^{[0]}\left(n_p^{[1]}+n_{-p}^{[1]}-1\right)} + 2S_p^2 C_p^2-(Z^{[0]}_p-4)S^2_pC^2_p\left(n_p+n_{-p}-1-2n_{p}n_{-p}\right)\right],
\end{align}
whereby, as expected, $\Tr{\rho_p^{[1|0]}}=1$.

We are now ready to compute the explicit expression of the average work $\mean{W[\Delta_1(\rho_0)]}$. This is achieved by following the same steps of the previous section and simply replacing $\rho_p^{[0]}$ with its diagonal part in $\rho_p^{[1|0]}$, basis of $H_{t_1}$. 
Accordingly,
\begin{align}
    \Tr{\rho^{[1|0]}_p{\Gamma_p^{[1]}}^\dagger \mathbb{M} \Gamma_p^{[1]}} &= \mathbb{M}_{11}\Tr{\rho_p^{[1|0]}n^{[1]}_p}+ \mathbb{M}_{22}\Tr{\rho^{[1|0]}_p\left(1-n_{-p}^{[1]}\right)} \\&= \frac{\mathbb{M}_{22}}{2} \tanh\left(\frac{\beta\omega_p^{[0]}}{2}\right)\left(C^4_p-S^4_p\right) = \frac{\mathbb{M}_{22}}{2} \tanh\left(\frac{\beta\omega_p^{[0]}}{2}\right)\cos(\phi_p^{[1]}-\phi_p^{[0]}) ,
\end{align}
where $\mathbb{M} =  R^\dagger_y(\phi_p^{[1]})\mathbb{U}_{p,t_{1}:t_{2}}^\dagger R^\dagger_y(\phi_p^{[2]}) \omega_p^{[2]}\sigma^z R_y(\phi_p^{[2]})\mathbb{U}_{p,t_{1}:t_{2}}R_y(\phi_p^{[1]})$. As a result, given that $\mathbb{M}_{22} = -\omega_p^{[2]}(2P^{[1,2]}_p-1)$ and $\cos(\phi^{[1]}_p-\phi^{[0]}_p) = 2P^{[0,1]}_p-1$, we finally get that
\begin{equation}
    \mean{W[\Delta_1(\rho_0)]} = \frac{L}{2\pi} \int_0^{\pi}dp\tanh\left(\frac{\beta\omega_p^{[0]}}{2}\right)\left[\omega_p^{[1]} \left(2P_p^{[0,1]}-1\right)-\omega_p^{[2]}\left(2P_p^{[1,2]}-1\right)\left(2P_p^{[0,1]}-1\right)\right].
\end{equation}
The formula above reduces to the expression for the average work shown in the main text upon substituting the parameters $Q_p^{[i,j]} = 2P_p^{[i,j]}-1$. 

\section*{Average work density and non-classical regions for different values of the model parameters}

\begin{figure}
    \centering
    \includegraphics[width=.9\linewidth]{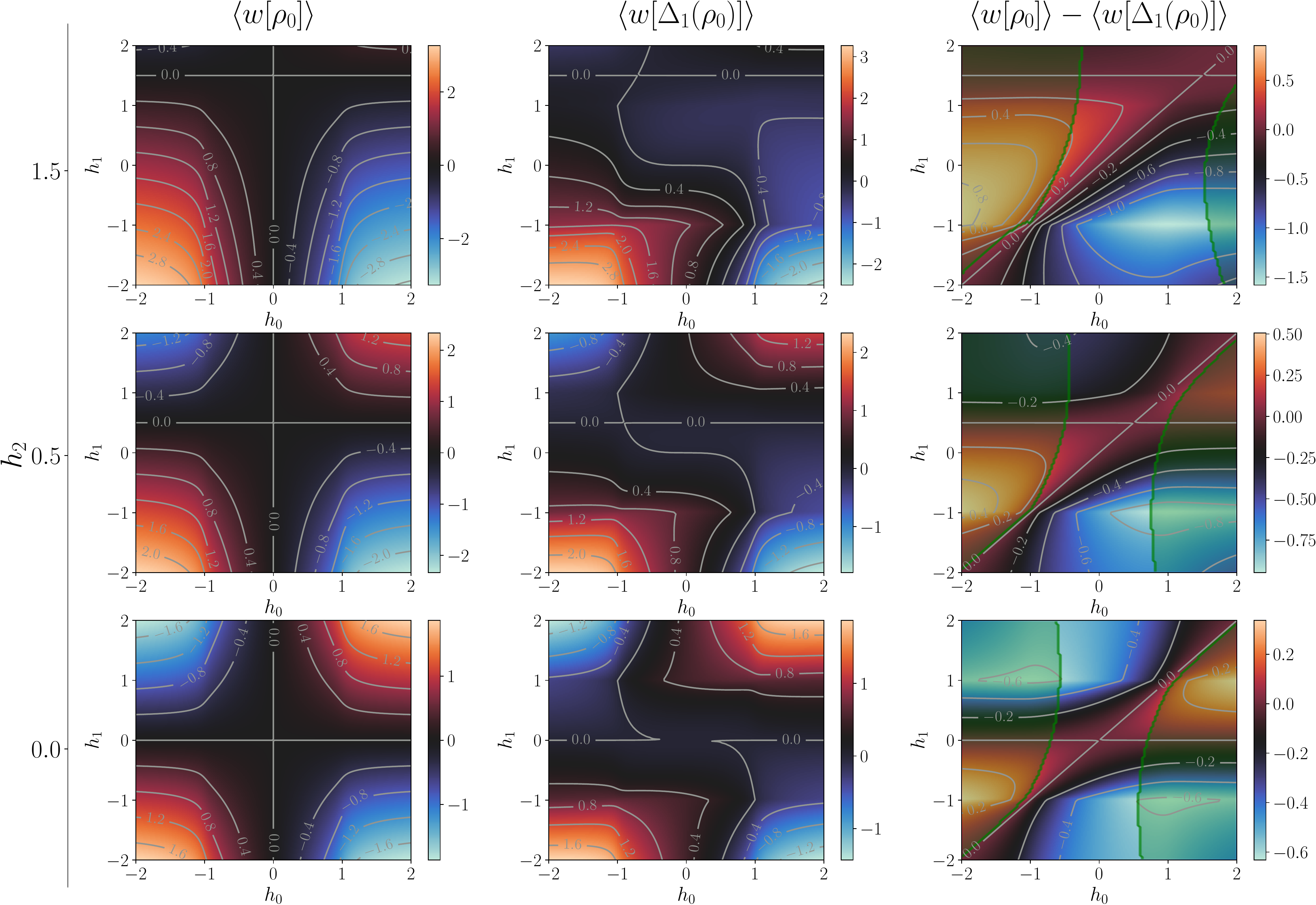}
    \caption{Average work density after a sudden quench from $h_1$ to $h_2$ for $\rho_0$, its dephased counterpart $\Delta_1(\rho_0)$ and the difference of the two. In all the plots $\beta = 15$.}
    \label{fig:Different_h2_SuddenQuench}
\end{figure}

For completeness, in Fig.~\ref{fig:Different_h2_SuddenQuench} we show the value of the average work density for $\rho_0$, its dephased counterpart $\Delta_1(\rho_0)$ and the difference of the two, with different choices of $h_2$. They are plotted by considering a sudden quench $h_1\to h_2$ and an inverse temperature of the initial quantum state equal to $\beta=15$. As in the main text, the shaded green areas mark the non-classicality regions, for which the fourth central moment of the KDQ distribution of work is negative. 

\begin{figure}
    \centering\includegraphics[width=.4\linewidth]{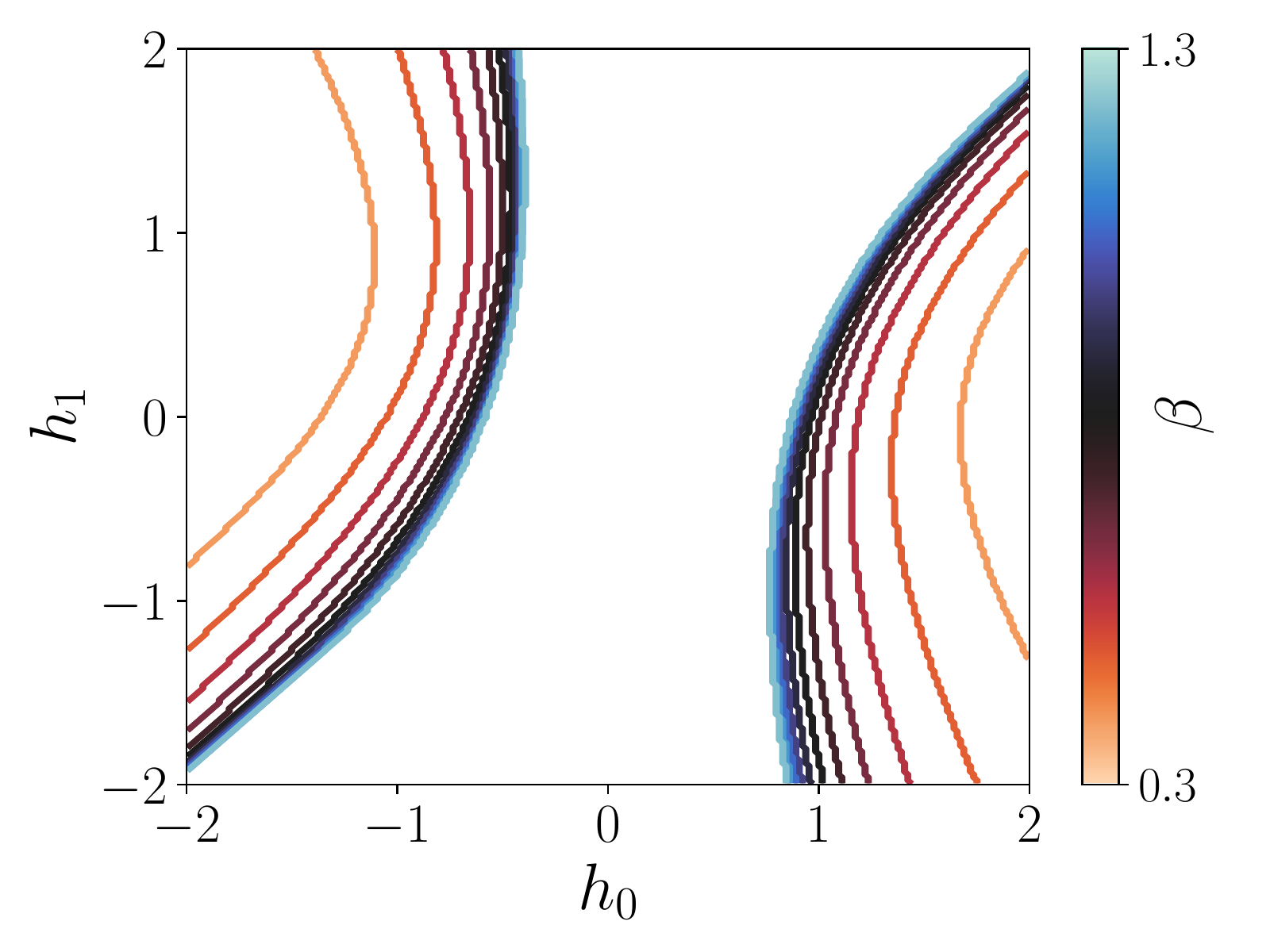}
    \caption{Quantum signatures of the KDQ distribution of work in the form of non-classical regions where the fourth central moment of the corresponding work distribution is negative. Such regions are obtained by varying the temperature of the initial state $\rho_0$, with a fixed value ($h_2 = 0.5$) of the final magnetic field. The border of the non-classical regions are marked with different colors corresponding to different values of temperature.}
    \label{fig:non_classicality_region_against_temperature}
\end{figure}

Moreover, in Fig.~\ref{fig:non_classicality_region_against_temperature}  we plot the border of the non-classicality region against $h_0$ and $h_1$ for a sudden quench with $h_2=0.5$. Different values of the initial inverse temperature $\beta$ are also considered. From the figure, it is worth noting that the non-classical region is reduced by increasing the temperature of the initial state. This signals that non-classical effects are reduced in the limit of high-temperature.

\section*{Relative entropy of quantum coherences}

\begin{figure}
    \centering
    \includegraphics[width=.7\linewidth]{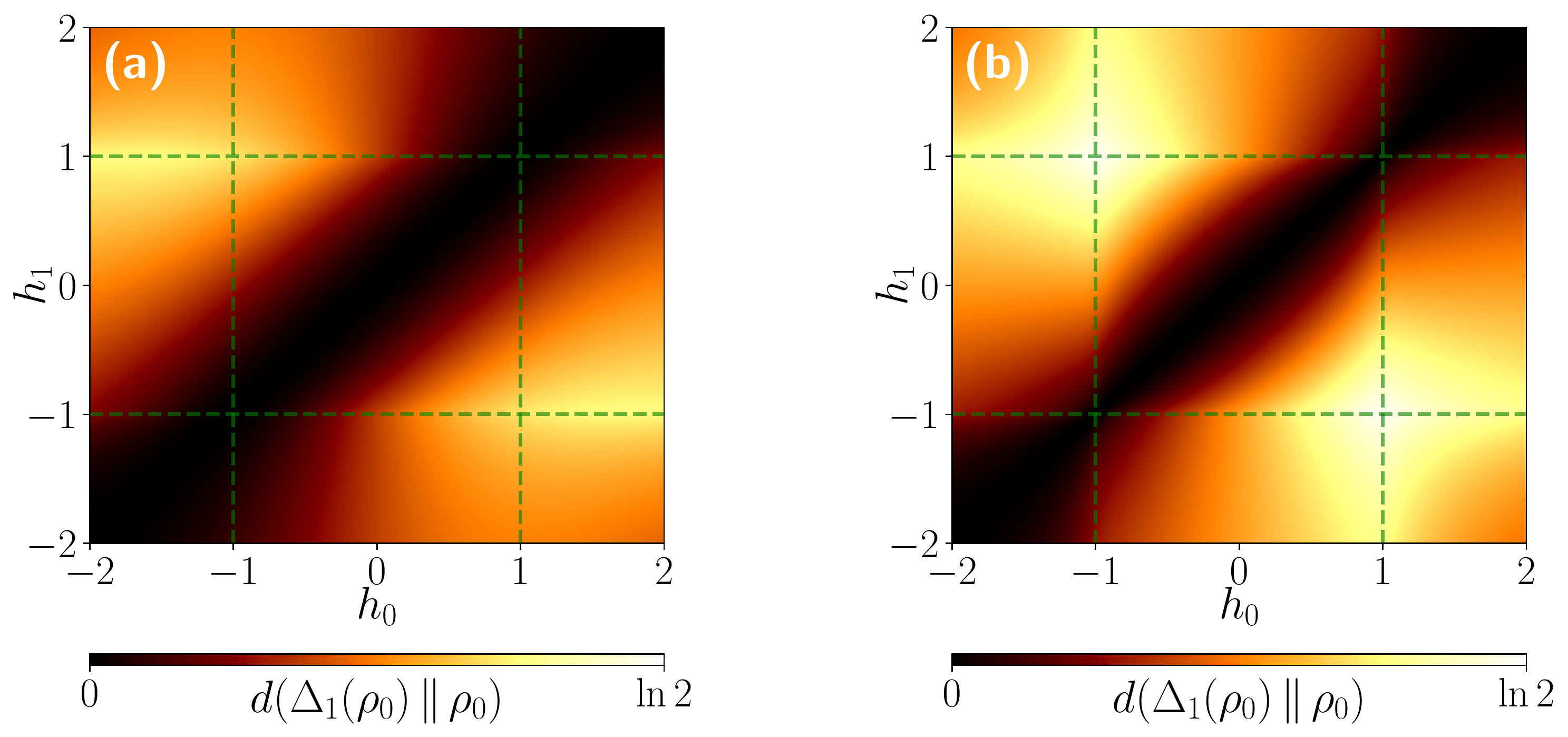}
    \caption{Relative entropy of the coherence density for the quantum Ising model between a thermal state $\rho_0=\exp{-\beta H_{t_0}}/Z$ with magnetic field $h=h_0$, and the dephased state $\Delta_1(\rho_0)$ with respect to the Hamiltonian $H_{t_1}$ with magnetic field $h=h_1$. Panel (a) $\beta=1$, Panel (b) $\beta=15$.}
\label{fig:RelativeEntropyOfCoherences}
\end{figure}

Let us consider the quantum relative entropy between the state $\rho_0$ and its projection on the basis of $H_{t_1}$. It is defined as
\begin{equation}
    D(\Delta_1(\rho_0) \,\|\, \rho_0) = \Tr{\rho_0\Big(\ln \rho_0 - \ln\Delta_1(\rho_0)\Big)}.
\end{equation}
It can be shown that
\begin{align}
     D(\Delta_1(\rho_0) \,\|\, \rho_0) = S[\Delta_1(\rho_0)]-S[\rho_0]
\end{align}
where $S[\rho]=-\Tr{\rho\ln\rho}$ is the von Neumann entropy of a quantum state $\rho$. We remind that $\rho_0 = \prod_p \rho_p^{[0]}$ and $\Delta_1(\rho_0) = \prod_p \rho_p^{[1|0]}$. Therefore, the quantum relative entropy among these two states can be also expressed by the following sum: 
\begin{equation}
    D(\Delta_1(\rho_0) \,\|\, \rho_0) = \sum_p \Tr{\rho_p^{[0]}\ln\rho_p^{[0]}-\rho_p^{[1|0]}\ln{\rho_p^{[1|0]}}}.
\end{equation}
Moreover, after some algebra, we also arrive at an explicit expression for the relative entropy of the coherence density:
\begin{align}
   \notag d(\Delta_1(\rho_0) \,\|\, \rho_0) = \frac{D(\Delta_1(\rho_0) \,\|\, \rho_0 )}{L} &= \frac{1}{\pi}\int_0^{\pi} dp \beta\omega_p^{[0]}\tanh{\frac{\beta\omega_p^{[0]}}{2}}+\notag\\
   -&\frac{1}{Z_p^{[0]}}\biggl[\left(P^{[1,0]}_p e^{\beta\omega_p^{[0]}}+(1-P^{[1,0]}_p) e^{-\beta\omega_p^{[0]}}\right)\ln{\left(P^{[1,0]}_p e^{\beta\omega_p^{[0]}}+(1-P^{[1,0]}_p) e^{-\beta\omega_p^{[0]}}\right)}+\notag\\
   +&\left(P^{[1,0]}_p e^{-\beta\omega_p^{[0]}}+(1-P^{[1,0]}_p) e^{\beta\omega_p^{[0]}}\right)\ln{\left(P^{[1,0]}_p e^{-\beta\omega_p^{[0]}}+(1-P^{[1,0]}_p) e^{\beta\omega_p^{[0]}}\right)} \biggr].
\end{align}
For the quantum Ising model, Fig.~\ref{fig:RelativeEntropyOfCoherences} shows $d(\Delta_1(\rho_0) \,\|\, \rho_0)$ as a function of $h_0$ and $h_1$, for two different values of the initial inverse temperature $\beta =1$ (panel a) and $\beta = 15$ (panel b). It is worth noting that $d(\Delta_1(\rho_0) \,\|\, \rho_0)$ has two maxima corresponding to $h_0 = \pm 1$ and $h_1 = \mp 1$, i.e., when $h_0$ and $h_1$ sit at the two quantum critical points of the model. At these points $P_p^{[0,1]} = 1/2$ ($Q_p^{[0,1]} = 0$) $\forall p$, thus leading to the maximum value 
\begin{align}
    \max_{h_0,h_1}d(\Delta_1(\rho_0) \,\|\, \rho_0) = \frac{1}{\pi}\int_0^\pi dp \left[\beta\omega_p^{[0]}\tanh\left( {\frac{\beta\omega_p^{[0]}}{2}} \right) -\frac{\cosh(\beta\omega_p^{[0]})}{1+\cosh(\beta\omega_p^{[0]})}\ln(\cosh(\beta\omega_p^{[0]}))\right].
\end{align}
Interestingly, 
\begin{align}
    \lim_{\beta\to 0}\max_{h_0,h_1}d(\Delta_1(\rho_0) \,\|\, \rho_0) = 0
\end{align} 
that signals again the fact that quantum coherences in $\rho_0$ (with respect to the basis of $H_{t_1}$) does not entail relevant effects to work statistics in the high-temperature limit. On the other hand, at low temperatures the relative entropy of quantum coherences increases, till to saturate the maximum value allowed by the Hilbert space dimension in the zero temperature (infinite $\beta$) limit:
\begin{align}
    \lim_{\beta\to \infty}\max_{h_0,h_1}d(\Delta_1(\rho_0) \,\|\, \rho_0) = \ln 2 \,.
\end{align}

\begin{figure}
    \centering
    \includegraphics[width=\linewidth]{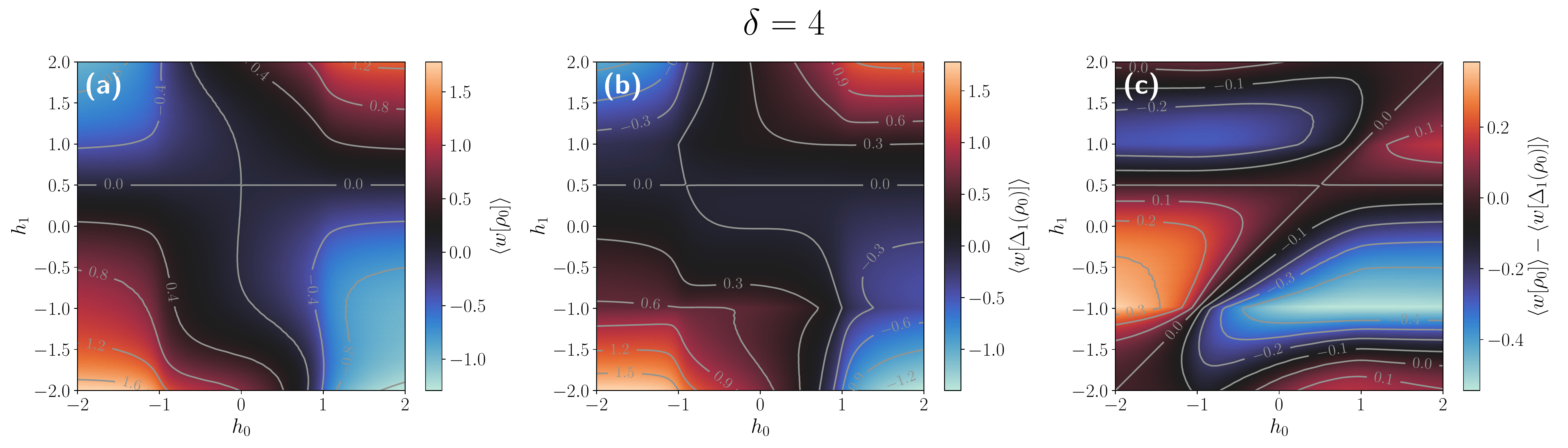}
    \includegraphics[width=\linewidth]{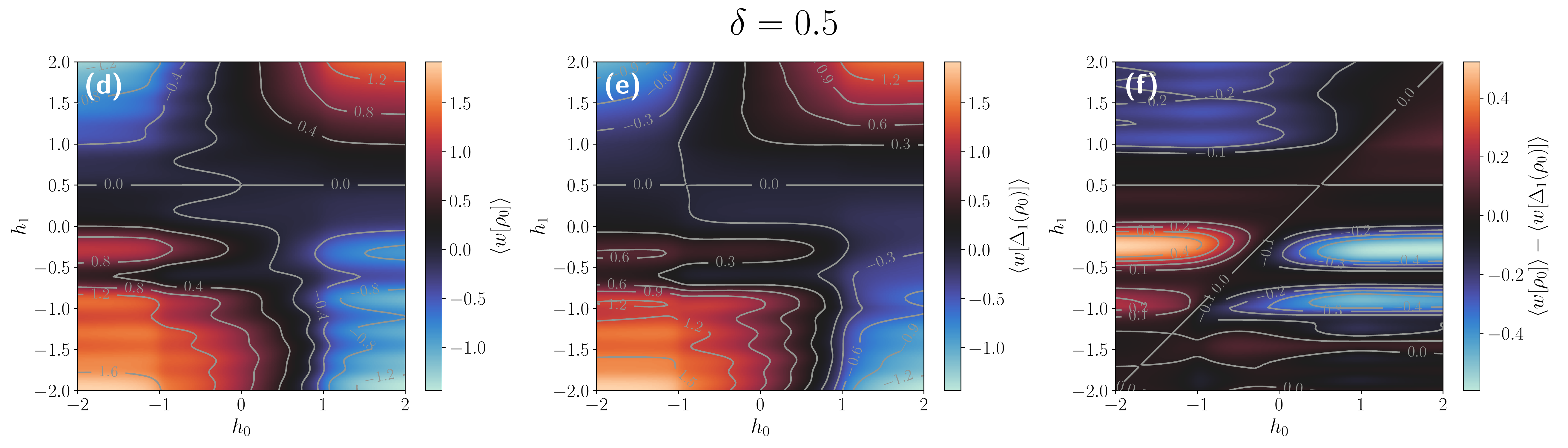}
    \caption{Average work density of the quantum Ising model obtained by initializing the system in the quantum state $\rho_0$ with magnetic field $h_0$ [panels (a) and (d)], or in the dephased state $\Delta_1(\rho_0)$ [panels (b) and (e)]. In all cases, the Hamiltonian changes are due to a linear driving of the magnetic field $h_1\to h_2$. Panels (c) and (f): Enhancement of work extraction. A faster driving allows for the extraction of a larger amount of work. (a),(b),(c) fast driving with velocity $\delta = 4$; (d),(e),(f) slow driving with velocity $\delta=0.5$.}
    \label{fig:TimeDependent quench}
\end{figure}

\section*{Landau-Zener-St{\"u}ckelberg-Majorana (LZSM) dynamics}

In order to properly take into account changes of the Hamiltonian at finite velocity, we have to solve the following time-dependent equation for $\mathbb{U}_{p,t_{1}:t_{2}}=
    \begin{pmatrix*}
        z_p(t) && -s^*_p(t) \\ s_p(t) && z^*_p(t)
    \end{pmatrix*}$:
\begin{equation}\label{eq: evoltuion equation}
    i\partial_t\begin{pmatrix*}[c]
        c_p(t)\\ c_{-p}^\dagger(t)
        \end{pmatrix*} =\left[(h(t)-\cos{p})\sigma^z-(\sin{p})\sigma^x\right]  \begin{pmatrix*}[c]
        c_p(t)\\ c_{-p}^\dagger(t)
        \end{pmatrix*}.
\end{equation}
The magnetic field $h(t)$ varies over time as a linear ramp at velocity $\delta$, i.e., 
\begin{equation}
    h(t) = h_1 + \delta (t-t_1)
\end{equation}
with $t\in\comm{t_1}{t_2}$ and $\delta = (h_2-h_1)/(t_2-t_1)$. The solution of Eq.\,\eqref{eq: evoltuion equation} can be determined by introducing the ansatz
\begin{equation}
    \begin{dcases}
        c_p(t) = z_p(t) c_p - s^*_p(t) c^\dagger_{-p}\\
        c_{-p}^\dagger(t) = s_p(t) c_p + z_p^*(t) c_{-p}^\dagger \,,
    \end{dcases}
\end{equation}
where the time dependence of the fermionic operators is encoded in the parameters $z_p(t)$ and $s_p(t)$. Therefore, the equation for $c_p(t)$ can be mapped into an equation for $z_p(t)$ and $s_p(t)$; in fact, from
\begin{equation}
    i\dot{z}_p(t) c_p - i\dot{s}^*_p(t) c^\dagger_{-p} = [h(t)-\cos{p}]\left(z_p(t) c_p - s^*_p(t) c^\dagger_{-p}\right)-\sin p \left[ s_p(t) c_p + z_p^*(t) c_{-p}^\dagger\right],
\end{equation}
we end up in the following equation in matrix form:
\begin{equation}\label{eq:LZSM_UnitaryOperator}
    i\begin{pmatrix*}[c]
        \dot{z}_p(t)\\ 
        \dot{s}_p(t)
    \end{pmatrix*} = \begin{pmatrix*}[c]
        \Omega_p(t) && \Delta_p\\ \Delta_p && -\Omega_p(t)
    \end{pmatrix*} \begin{pmatrix*}
        z_p(t) \\ s_p(t)
    \end{pmatrix*} = \mathbb{H}_p(t)\begin{pmatrix*}
        z_p(t) \\ s_p(t)
    \end{pmatrix*},
\end{equation}
where $\Omega_p(t) = h(t)-\cos{p}$ and $\Delta_p=-\sin{p}$, with initial conditions $z_p(t_1)=1$ and $s_p(t_1)=0$. The first order equations for $z_p(t_1)=1$ and $s_p(t_1)=0$ are not independent and can be thus recast into a second order equation for $z_p(t)$, i.e.,
\begin{equation}
    i \ddot{z}_p=\dot{\Omega}_p z_p + \Omega_p \dot{z}_p + \Delta_p \dot{s}_p = \delta z_p -i\Omega_p\left[\Omega_p z_p+\Delta_p s_p\right]-i\Delta_p\left[\Delta_p z_p - \Omega_p s_p\right]
\end{equation}
that reduces to
\begin{equation}
    \ddot{z}_p + \left[\Omega_p^2(t)+\Delta^2_p+i\delta\right]z_p = 0 \,.
\end{equation}
This equation has to be solved with initial conditions $z_p(t_1) = 1$ and $\dot{z}_p =-i\Omega_p(t_1)$, while 
\begin{equation}
    s_p(t) = \frac{1}{\Delta_p}\left[i\partial_t -\Omega_p(t)\right]z_p(t) \,.
\end{equation}
The general solution for $\mathbb{U}_{p,t_{1}:t_{2}}$ is readily available in terms of the Weber D-functions~\cite{Vitanov1996PRA}. From this solution we can compute the exact transition probabilities, as well as $Q_p^{[0,2]}$. Moreover, from the general formulas in the main text about quadratic fermionic models, also the explicit average work originated by a linear driving can be determined. We plot in Fig.~\ref{fig:TimeDependent quench} the average work density that is obtained by initializing the system in the quantum state $\rho_0$ or in its dephased counterpart $\Delta_1(\rho_0)$. In doing this, we plot their difference both for a fast ($\delta=4$) and slow driving ($\delta = 0.5$). It is worth noting that the average work density $\mean{w[\rho_0]}$ witnesses an enhancement of the work extraction, which is boosted by the non-commutativity of $\rho_0$ with $H_{t_1}$. The latter makes available to us quantum coherence of the initial quantum state $\rho_0$ that is {\it robust} against finite but sufficiently fast velocity of the driving. Moreover, also observe that in the opposite limit of an infinitely slow driving, the difference $\mean{w[\rho_0]} -\mean{w[\Delta_1(\rho_0)]}$ approaches zero whenever the driving ramp does not cross a quantum critical point, while it takes a finite value around the criticality (see Fig. \ref{fig:TimeDependent quench}f). This may be ascribed to the fact that, in case the adiabatic hypothesis is satisfied, $Q_p^{[1,2]}\to 1$ and $Q_p^{[0,2]}\to Q_p^{[0,1]}$ as $\delta\to 0$, thus leading to a zero average work difference. However, the adiabatic hypothesis is violated whenever a gapless quantum critical point is crossed during a quantum dynamics. Such violation allows for a finite enhancement of extracted work that persists also for slow drivings.

\end{widetext}

\end{document}